\input epsf

\catcode`\@=11 
%

\font\fourteenrm=cmr10 scaled\magstep2
\font\twelverm=cmr10 scaled\magstep1
\font\elevenrm=cmr10 scaled\magstephalf
\font\ninerm=cmr9	     \font\sixrm=cmr6
\font\seventeenbf=cmbx10 scaled\magstep3
\font\fourteenbf=cmbx10 scaled\magstep2
\font\twelvebf=cmbx10 scaled\magstep1
\font\elevenbf=cmbx10 scaled\magstephalf
\font\ninebf=cmbx9	      \font\sixbf=cmbx6
\font\seventeeni=cmmi10 scaled\magstep3	    \skewchar\seventeeni='177
\font\fourteeni=cmmi10 scaled\magstep2	    \skewchar\fourteeni='177
\font\twelvei=cmmi10 scaled\magstep1	    \skewchar\twelvei='177
\font\eleveni=cmmi10 scaled\magstephalf \skewchar\eleveni='177
\font\ninei=cmmi9			    \skewchar\ninei='177
\font\sixi=cmmi6			    \skewchar\sixi='177
\font\seventeensy=cmsy10 scaled\magstep3    \skewchar\seventeensy='60
\font\fourteensy=cmsy10 scaled\magstep2	    \skewchar\fourteensy='60
\font\twelvesy=cmsy10 scaled\magstep1	    \skewchar\twelvesy='60
\font\elevensy=cmsy10 scaled\magstephalf \skewchar\elevensy='60
\font\ninesy=cmsy9			    \skewchar\ninesy='60
\font\sixsy=cmsy6			    \skewchar\sixsy='60

\font\fourteenex=cmex10 scaled\magstep2
\font\twelveex=cmex10 scaled\magstep1
\font\elevenex=cmex10 scaled\magstephalf

\font\fourteensl=cmsl10 scaled\magstep2
\font\twelvesl=cmsl10 scaled\magstep1
\font\elevensl=cmsl10 scaled\magstephalf
\font\ninesl=cmsl9

\font\fourteenit=cmti10 scaled\magstep2
\font\twelveit=cmti10 scaled\magstep1
\font\elevenit=cmti10 scaled\magstephalf
\font\twelvett=cmtt10 scaled\magstep1
\font\eleventt=cmtt10 scaled\magstephalf
\font\twelvecp=cmcsc10 scaled\magstep1
\font\elevencp=cmcsc10 scaled\magstephalf
\font\tencp=cmcsc10
\newfam\cpfam
\newcount\f@ntkey	     \f@ntkey=0
\def\samef@nt{\relax \ifcase\f@ntkey \rm \or\oldstyle \or\or
	 \or\it \or\sl \or\bf \or\tt \or\caps \fi }
\def\fourteenpoint{\relax
   \textfont0=\fourteenrm	    \scriptfont0=\tenrm
   \scriptscriptfont0=\sevenrm
   \def\rm{\fam0 \fourteenrm \f@ntkey=0 }%
   \textfont1=\fourteeni	    \scriptfont1=\teni
   \scriptscriptfont1=\seveni
   \def\oldstyle{\fam1 \fourteeni\f@ntkey=1 }%
   \textfont2=\fourteensy	    \scriptfont2=\tensy
   \scriptscriptfont2=\sevensy
   \textfont3=\fourteenex     \scriptfont3=\fourteenex
   \scriptscriptfont3=\fourteenex
   \def\it{\fam\itfam \fourteenit\f@ntkey=4 }\textfont\itfam=\fourteenit
   \def\sl{\fam\slfam \fourteensl\f@ntkey=5 }\textfont\slfam=\fourteensl
   \scriptfont\slfam=\tensl
   \def\bf{\fam\bffam \fourteenbf\f@ntkey=6 }\textfont\bffam=\fourteenbf
   \scriptfont\bffam=\tenbf	 \scriptscriptfont\bffam=\sevenbf
   \def\tt{\fam\ttfam \twelvett \f@ntkey=7 }\textfont\ttfam=\twelvett
   \h@big=11.9\p@ \h@Big=16.1\p@ \h@bigg=20.3\p@ \h@Bigg=24.5\p@
   \def\caps{\fam\cpfam \twelvecp \f@ntkey=8 }\textfont\cpfam=\twelvecp
   \setbox\strutbox=\hbox{\vrule height 12pt depth 5pt width\z@}%
   \samef@nt}
\def\twelvepoint{\relax
   \textfont0=\twelverm	  \scriptfont0=\ninerm
   \scriptscriptfont0=\sevenrm
   \def\rm{\fam0 \twelverm \f@ntkey=0 }%
   \textfont1=\twelvei		  \scriptfont1=\ninei
   \scriptscriptfont1=\seveni
   \def\oldstyle{\fam1 \twelvei\f@ntkey=1 }%
   \textfont2=\twelvesy	  \scriptfont2=\ninesy
   \scriptscriptfont2=\sevensy
   \textfont3=\twelveex	  \scriptfont3=\twelveex
   \scriptscriptfont3=\twelveex
   \def\it{\fam\itfam \twelveit \f@ntkey=4 }\textfont\itfam=\twelveit
   \def\sl{\fam\slfam \twelvesl \f@ntkey=5 }\textfont\slfam=\twelvesl
   \scriptfont\slfam=\ninesl
   \def\bf{\fam\bffam \twelvebf \f@ntkey=6 }\textfont\bffam=\twelvebf
   \scriptfont\bffam=\ninebf	  \scriptscriptfont\bffam=\sevenbf
   \def\tt{\fam\ttfam \twelvett \f@ntkey=7 }\textfont\ttfam=\twelvett
   \h@big=10.2\p@ \h@Big=13.8\p@ \h@bigg=17.4\p@ \h@Bigg=21.0\p@ 
   \def\caps{\fam\cpfam \twelvecp \f@ntkey=8 }\textfont\cpfam=\twelvecp
   \setbox\strutbox=\hbox{\vrule height 10pt depth 4pt width\z@}%
   \samef@nt}
\def\elevenpoint{\relax
 \textfont0=\elevenrm \scriptfont0=\ninerm \scriptscriptfont0=\sixrm
 \def\rm{\fam0 \elevenrm \f@ntkey=0}\relax
 \textfont1=\eleveni \scriptfont1=\ninei \scriptscriptfont1=\sixi
 \def\oldstyle{\fam1 \eleveni\f@ntkey=1}\relax
 \textfont2=\elevensy \scriptfont2=\ninesy \scriptscriptfont2=\sixsy
 \textfont3=\elevenex \scriptfont3=\elevenex \scriptscriptfont3=\elevenex
 \def\it{\fam\itfam \elevenit \f@ntkey=4 }\textfont\itfam=\elevenit
 \def\sl{\fam\slfam \elevensl \f@ntkey=5 }\textfont\slfam=\elevensl
 \scriptfont\slfam=\ninesl
 \def\bf{\fam\bffam \elevenbf \f@ntkey=6 }\textfont\bffam=\elevenbf
 \scriptfont\bffam=\ninebf \scriptscriptfont\bffam=\sixbf
 \def\tt{\fam\ttfam \eleventt \f@ntkey=7 }\textfont\ttfam=\eleventt
 \h@big=9.311\p@ \h@Big=12.6\p@ \h@bigg=15.88\p@ \h@Bigg=19.17\p@
 \def\caps{\fam\cpfam \elevencp \f@ntkey=8 }\textfont\cpfam=\elevencp
 \setbox\strutbox=\hbox{\vrule height 9pt depth 4pt width\z@}\relax
 \samef@nt}
\def\tenpoint{\relax
   \textfont0=\tenrm	       \scriptfont0=\sevenrm
   \scriptscriptfont0=\fiverm
   \def\rm{\fam0 \tenrm \f@ntkey=0 }%
   \textfont1=\teni	       \scriptfont1=\seveni
   \scriptscriptfont1=\fivei
   \def\oldstyle{\fam1 \teni \f@ntkey=1 }%
   \textfont2=\tensy	       \scriptfont2=\sevensy
   \scriptscriptfont2=\fivesy
   \textfont3=\tenex	       \scriptfont3=\tenex
   \scriptscriptfont3=\tenex
   \def\it{\fam\itfam \tenit \f@ntkey=4 }\textfont\itfam=\tenit
   \def\sl{\fam\slfam \tensl \f@ntkey=5 }\textfont\slfam=\tensl
   \def\bf{\fam\bffam \tenbf \f@ntkey=6 }\textfont\bffam=\tenbf
   \scriptfont\bffam=\sevenbf	   \scriptscriptfont\bffam=\fivebf
   \def\tt{\fam\ttfam \tentt \f@ntkey=7 }\textfont\ttfam=\tentt
   \def\caps{\fam\cpfam \tencp \f@ntkey=8 }\textfont\cpfam=\tencp
   \h@big=8.5\p@ \h@Big=11.5\p@ \h@bigg=14.5\p@ \h@Bigg=17.5\p@ 
   \setbox\strutbox=\hbox{\vrule height 8.5pt depth 3.5pt width\z@}%
   \samef@nt}
\newdimen\h@big  \h@big=8.5\p@
\newdimen\h@Big  \h@Big=11.5\p@
\newdimen\h@bigg  \h@bigg=14.5\p@
\newdimen\h@Bigg  \h@Bigg=17.5\p@
\def\big#1{{\hbox{$\left#1\vbox to\h@big{}\right.\n@space$}}}
\def\Big#1{{\hbox{$\left#1\vbox to\h@Big{}\right.\n@space$}}}
\def\bigg#1{{\hbox{$\left#1\vbox to\h@bigg{}\right.\n@space$}}}
\def\Bigg#1{{\hbox{$\left#1\vbox to\h@Bigg{}\right.\n@space$}}}
\normalbaselineskip = 20pt plus 0.2pt minus 0.1pt
\normallineskip = 1.5pt plus 0.1pt minus 0.1pt
\normallineskiplimit = 1.5pt
\newskip\normaldisplayskip
\normaldisplayskip = 20pt plus 5pt minus 10pt
\newskip\normaldispshortskip
\normaldispshortskip = 6pt plus 5pt
\newskip\normalparskip
\normalparskip = 6pt plus 2pt minus 1pt
\newskip\skipregister
\skipregister = 5pt plus 2pt minus 1.5pt
\newif\ifsingl@	   \newif\ifdoubl@
\newif\iftwelv@	   \twelv@true
\newif\ifelev@n \elev@nfalse
\def\singlespace{\singl@true\doubl@false\spaces@t}
\def\doublespace{\singl@false\doubl@true\spaces@t}
\def\normalspace{\singl@false\doubl@false\spaces@t}
\def\Tenpoint{\tenpoint\twelv@false\elev@nfalse\spaces@t}
\def\Elevenpoint{\elevenpoint\twelv@false\elev@ntrue\spaces@t}
\def\Twelvepoint{\twelvepoint\twelv@true\elev@nfalse\spaces@t}
\def\spaces@t{\relax
\iftwelv@ \ifsingl@\subspaces@t3:4;\else\subspaces@t1:1;\fi
\else \ifelev@n \ifsingl@\subspaces@t2:3;\else\subspaces@t9:10;\fi
\else \ifsingl@\subspaces@t3:5;\else\subspaces@t4:5;\fi \fi \fi
\ifdoubl@ \multiply\baselineskip by 5
\divide\baselineskip by 4 \fi}
\def\subspaces@t#1:#2;{\baselineskip=\normalbaselineskip
   \multiply\baselineskip by #1\divide\baselineskip by #2%
   \lineskip = \normallineskip
   \multiply\lineskip by #1\divide\lineskip by #2%
   \lineskiplimit = \normallineskiplimit
   \multiply\lineskiplimit by #1\divide\lineskiplimit by #2%
   \parskip = \normalparskip
   \multiply\parskip by #1\divide\parskip by #2%
   \abovedisplayskip = \normaldisplayskip
   \multiply\abovedisplayskip by #1\divide\abovedisplayskip by #2%
   \belowdisplayskip = \abovedisplayskip
   \abovedisplayshortskip = \normaldispshortskip
   \multiply\abovedisplayshortskip by #1%
   \divide\abovedisplayshortskip by #2%
   \belowdisplayshortskip = \abovedisplayshortskip
   \advance\belowdisplayshortskip by \belowdisplayskip
   \divide\belowdisplayshortskip by 2
   \smallskipamount = \skipregister
   \multiply\smallskipamount by #1\divide\smallskipamount by #2%
   \medskipamount = \smallskipamount \multiply\medskipamount by 2 
   \bigskipamount = \smallskipamount \multiply\bigskipamount by 4 }
\def\normalbaselines{ \baselineskip=\normalbaselineskip%
   \lineskip=\normallineskip \lineskiplimit=\normallineskip%
   \iftwelv@\else \multiply\baselineskip by 4 \divide\baselineskip by 5%
   \multiply\lineskiplimit by 4 \divide\lineskiplimit by 5%
   \multiply\lineskip by 4 \divide\lineskip by 5 \fi }
\Twelvepoint  
\interlinepenalty=50
\interfootnotelinepenalty=5000
\predisplaypenalty=9000
\postdisplaypenalty=500
\hfuzz=1pt
\vfuzz=0.2pt
%
%
%
\def\pagecontents{%
   \ifvoid\topins\else\unvbox\topins\vskip\skip\topins\fi
   \dimen@ = \dp255 \unvbox255
   \ifvoid\footins\else\vskip\skip\footins\footrule\unvbox\footins\fi
   \ifr@ggedbottom \kern-\dimen@ \vfil \fi }
\def\makeheadline{\vbox to 0pt{ \skip@=\topskip
   \advance\skip@ by -12pt \advance\skip@ by -2\normalbaselineskip
   \vskip\skip@ \line{\vbox to 12pt{}\the\headline} \vss
   }\nointerlineskip}
\def\makefootline{\baselineskip = 1.5\normalbaselineskip
   \line{\the\footline}}
\newif\iffrontpage
\newif\ifp@genum
\def\nopagenumbers{\p@genumfalse}
\def\pagenumbers{\p@genumtrue}
\pagenumbers
\newtoks\date
\newtoks\Month
\footline={\hss\iffrontpage\else\ifp@genum\tenrm\folio\hss\fi\fi}
\headline={\iffinal\hfil\else\tenrm DRAFT\hfil\the\date\fi}
\def\monthname{\relax\ifcase\month 0/\or January\or February\or
   March\or April\or May\or June\or July\or August\or September\or
   October\or November\or December\else\number\month/\fi}
\date={\monthname\ \number\day, \number\year}
\Month={\monthname\ \number\year}
\countdef\pagenumber=1  \pagenumber=1
\def\advancepageno{\global\advance\pageno by 1
   \ifnum\pagenumber<0 \global\advance\pagenumber by -1
   \else\global\advance\pagenumber by 1 \fi \global\frontpagefalse }
\def\folio{\ifnum\pagenumber<0 \romannumeral-\pagenumber
   \else \number\pagenumber \fi }
\def\footrule{\dimen@=\prevdepth\nointerlineskip
   \vbox to 0pt{\vskip -0.25\baselineskip \hrule width 0.35\hsize \vss}
   \prevdepth=\dimen@ }
\newtoks\foottokens
\foottokens={\Tenpoint\singlespace}
\newdimen\footindent
\footindent=24pt
\def\vfootnote#1{\insert\footins\bgroup  \the\foottokens
   \interlinepenalty=\interfootnotelinepenalty \floatingpenalty=20000
   \splittopskip=\ht\strutbox \boxmaxdepth=\dp\strutbox
   \leftskip=\footindent \rightskip=\z@skip
   \parindent=0.5\footindent \parfillskip=0pt plus 1fil
   \spaceskip=\z@skip \xspaceskip=\z@skip
   \Textindent{$ #1 $}\footstrut\futurelet\next\fo@t}
\def\Textindent#1{\noindent\llap{#1\enspace}\ignorespaces}
\def\footnote#1{\attach{#1}\vfootnote{#1}}

\let\footsymbol=\star
\newcount\lastf@@t	     \lastf@@t=-1
\newcount\footsymbolcount    \footsymbolcount=0
\newif\ifPhysRev
\def\footsymbolgen{\relax \ifPhysRev \iffrontpage \NPsymbolgen\else
   \PRsymbolgen\fi \else \NPsymbolgen\fi
   \global\lastf@@t=\pageno \footsymbol }
\def\NPsymbolgen{\ifnum\footsymbolcount<0 \global\footsymbolcount=0\fi
   {\iffrontpage \else \advance\lastf@@t by 1 \fi
   \ifnum\lastf@@t<\pageno \global\footsymbolcount=0
   \else \global\advance\footsymbolcount by 1 \fi }
   \ifcase\footsymbolcount \fd@f\star\or \fd@f\dagger\or \fd@f\ast\or
   \fd@f\ddagger\or \fd@f\natural\or \fd@f\diamond\or \fd@f\bullet\or
   \fd@f\nabla\else \fd@f\dagger\global\footsymbolcount=0 \fi }
\def\fd@f#1{\xdef\footsymbol{#1}}
\def\PRsymbolgen{\ifnum\footsymbolcount>0 \global\footsymbolcount=0\fi
   \global\advance\footsymbolcount by -1
   \xdef\footsymbol{\sharp\number-\footsymbolcount} }
\def\space@ver#1{\let\@sf=\empty \ifmmode #1\else \ifhmode
   \edef\@sf{\spacefactor=\the\spacefactor}\unskip${}#1$\relax\fi\fi}
\def\attach#1{\space@ver{\strut^{\mkern 2mu #1} }\@sf\ }
%
%
%
\newcount\chapternumber	     \chapternumber=0
\newcount\sectionnumber	     \sectionnumber=0
\newcount\equanumber	     \equanumber=0
\let\chapterlabel=\relax
\newtoks\chapterstyle	     \chapterstyle={\Number}
\newskip\chapterskip	     \chapterskip=\bigskipamount
\newskip\sectionskip	     \sectionskip=\medskipamount
\newskip\headskip	     \headskip=8pt plus 3pt minus 3pt
\newdimen\chapterminspace    \chapterminspace=15pc
\newdimen\sectionminspace    \sectionminspace=10pc
\newdimen\referenceminspace  \referenceminspace=25pc
\def\chapterreset{\global\advance\chapternumber by 1
   \ifnum\equanumber<0 \else\global\equanumber=0\fi
   \sectionnumber=0 \makel@bel}
\def\makel@bel{\xdef\chapterlabel{%
   \the\chapterstyle{\the\chapternumber}.}}
\def\sectionlabel{\number\sectionnumber \quad }
\def\alphabetic#1{\count255='140 \advance\count255 by #1\char\count255}
\def\Alphabetic#1{\count255='100 \advance\count255 by #1\char\count255}
\def\Roman#1{\uppercase\expandafter{\romannumeral #1}}
\def\roman#1{\romannumeral #1}
\def\Number#1{\number #1}
\def\unnumberedchapters{\let\makel@bel=\relax \let\chapterlabel=\relax
\let\sectionlabel=\relax \equanumber=-1 }
\def\titlestyle#1{\par\begingroup \interlinepenalty=9999
   \leftskip=0.02\hsize plus 0.23\hsize minus 0.02\hsize
   \rightskip=\leftskip \parfillskip=0pt
   \hyphenpenalty=9000 \exhyphenpenalty=9000
   \tolerance=9999 \pretolerance=9000
   \spaceskip=0.333em \xspaceskip=0.5em
   \iftwelv@\fourteenpoint\else\twelvepoint\fi
   \noindent #1\par\endgroup }
\def\spacecheck#1{\dimen@=\pagegoal\advance\dimen@ by -\pagetotal
   \ifdim\dimen@<#1 \ifdim\dimen@>0pt \vfil\break \fi\fi}
\def\chapter#1{\par \penalty-300 \vskip\chapterskip
   \spacecheck\chapterminspace
   \chapterreset \titlestyle{\chapterlabel \ #1}
   \nobreak\vskip\headskip \penalty 30000
   \wlog{\string\chapter\ \chapterlabel} }

\def\section#1{\par \ifnum\the\lastpenalty=30000\else
   \penalty-200\vskip\sectionskip \spacecheck\sectionminspace\fi
   \wlog{\string\section\ \chapterlabel \the\sectionnumber}
   \global\advance\sectionnumber by 1  \noindent
   {\caps\enspace\chapterlabel \sectionlabel #1}\par
   \nobreak\vskip\headskip \penalty 30000 }
\def\subsection#1{\par
   \ifnum\the\lastpenalty=30000\else \penalty-100\smallskip \fi
   \noindent\undertext{#1}\enspace \vadjust{\penalty5000}}

\def\undertext#1{\vtop{\hbox{#1}\kern 1pt \hrule}}
\def\ack{\par\penalty-100\medskip \spacecheck\sectionminspace
   \line{\fourteenrm\hfil ACKNOWLEDGMENTS\hfil}\nobreak\vskip\headskip }
%
\def\APPENDIX#1#2{\par\penalty-300\vskip\chapterskip
   \spacecheck\chapterminspace \chapterreset \xdef\chapterlabel{#1}
   \titlestyle{APPENDIX #2} \nobreak\vskip\headskip \penalty 30000
   \wlog{\string\Appendix\ \chapterlabel} }
\def\Appendix#1{\APPENDIX{#1}{#1}}
\def\appendix{\APPENDIX{A}{}}
%
%
%
\newif\iffinal \finaltrue
\def\showeqname#1{\iffinal\else\hbox to 0pt{\tentt\kern2mm\string#1\hss}\fi}
\def\showEqname#1{\iffinal\else \hskip 0pt plus 1fill
 \hbox to 0pt{\tentt\kern2mm\string#1\hss}\hskip 0pt plus -1fill\fi}
\def\eqnamedef#1{\relax \ifnum\equanumber<0
   \xdef#1{{\noexpand\rm(\number-\equanumber)}}\global\advance\equanumber by -1
   \else \global\advance\equanumber by 1
   \xdef#1{{\noexpand\rm(\chapterlabel \number\equanumber)}}\fi}
\def\eqnamenewdef#1#2{\relax \ifnum\equanumber<0
   \xdef#1{{\noexpand\rm(\number-\equanumber#2)}}\global\advance\equanumber 
   by -1 \else \global\advance\equanumber by 1
   \xdef#1{{\noexpand\rm(\chapterlabel \number\equanumber#2)}}\fi}
\def\eqnameolddef#1#2{\relax \ifnum\equanumber<0
   \global\advance\equanumber by 1 
   \xdef#1{{\noexpand\rm(\number-\equanumber#2)}}\global\advance\equanumber 
   by -1 \else \xdef#1{{\noexpand\rm(\chapterlabel \number\equanumber#2)}}\fi}
\def\eqname#1{\eqnamedef{#1}#1}
\def\eqnamenew#1#2{\eqnamenewdef{#1}{#2}#1}
\def\eqnameold#1#2{\eqnameolddef{#1}{#2}#1}
\def\eq{\eqname\lasteq}
\def\eqa{\eqnamenew\lasteq a}
\def\eqb{\eqnameold\lasteq b}
\def\eqc{\eqnameold\lasteq c}
\def\eqd{\eqnameold\lasteq d}
\def\eqnew#1{\eqnamenew\lasteq{#1}}
\def\eqold#1{\eqnameold\lasteq{#1}}
\def\eq@@{\ifinner\let\eqn@=\relax\else\let\eqn@=\eqno\fi\eqn@}
\def\Eq{\eq@@\eq}
\def\Eqnew#1{\eq@@\eqnew{#1}}
\def\Eqold#1{\eq@@\eqold{#1}}
\def\Eqa{\eq@@\eqa}
\def\Eqb{\eq@@\eqb}
\def\Eqc{\eq@@\eqc}
\def\Eqd{\eq@@\eqd}
\def\Eqn#1{\eq@@\eqname{#1}\showeqname{#1}}
\def\Eqnnew#1#2{\eq@@\eqnamenew{#2}{#1}\showeqname{#1}}
\def\Eqnold#1#2{\eq@@\eqnameold{#2}{#1}\showeqname{#1}}
\def\Eqna#1{\eq@@\eqnamenew{#1}a\showeqname{#1}}
\def\Eqnb#1{\eq@@\eqnameold{#1}b\showeqname{#1}}
\def\Eqnc#1{\eq@@\eqnameold{#1}c\showeqname{#1}}
\def\Eqnd#1{\eq@@\eqnameold{#1}d\showeqname{#1}}

%

%
%
\def\GENITEM#1;#2{\par \hangafter=0 \hangindent=#1
   \Textindent{$ #2 $}\ignorespaces}
\outer\def\newitem#1=#2;{\gdef#1{\GENITEM #2;}}
\newdimen\itemsize		  \itemsize=30pt
\newitem\item=1\itemsize;
\newitem\sitem=1.75\itemsize;	  
\newitem\ssitem=2.5\itemsize;	  
\outer\def\newlist#1=#2&#3&#4;{\toks0={#2}\toks1={#3}%
   \count255=\escapechar \escapechar=-1
   \alloc@0\list\countdef\insc@unt\listcount	 \listcount=0
   \edef#1{\par
      \countdef\listcount=\the\allocationnumber
      \advance\listcount by 1
      \hangafter=0 \hangindent=#4
      \Textindent{\the\toks0{\listcount}\the\toks1}}
   \expandafter\expandafter\expandafter
   \edef\c@t#1{begin}{\par
      \countdef\listcount=\the\allocationnumber \listcount=1
      \hangafter=0 \hangindent=#4
      \Textindent{\the\toks0{\listcount}\the\toks1}}
   \expandafter\expandafter\expandafter
   \edef\c@t#1{con}{\par \hangafter=0 \hangindent=#4 \noindent}
   \escapechar=\count255}
\def\c@t#1#2{\csname\string#1#2\endcsname}
\newlist\point=\Number&.&1.0\itemsize;
\newlist\subpoint=(\alphabetic&)&1.75\itemsize;
\newlist\subsubpoint=(\roman&)&2.5\itemsize;
%

%
%
%
\def\keepspacefactor{\let\@sf=\empty \ifhmode
   \edef\@sf{\spacefactor=\the\spacefactor\relax}\relax\fi}
\newcount\footcount \footcount=0
\def\Footnote{\global\advance\footcount by 1 \footnote{\the\footcount}}
\def\footnote#1{\keepspacefactor\refattach{#1}\vfootnote{#1}}

\def\nonfrenchspacing{\sfcode\lq\.=3000 \sfcode\lq\?=3001 \sfcode\lq\!=3001
 \sfcode\lq\:=2000 \sfcode\lq\;=1500 \sfcode\lq\,=1250 }

\nonfrenchspacing
\newcount\referencecount     \referencecount=0
\newif\ifreferenceopen	     \newwrite\referencewrite
\newtoks\rw@toks
\newcount\lastrefsbegincount \lastrefsbegincount=0
\def\refsend{\refmark{\count255=\referencecount
   \advance\count255 by-\lastrefsbegincount
   \ifcase\count255 \number\referencecount
   \or \number\lastrefsbegincount,\number\referencecount
   \else \number\lastrefsbegincount-\number\referencecount \fi}}
\def\refch@ck{\chardef\rw@write=\referencewrite
   \ifreferenceopen \else \referenceopentrue
   \immediate\openout\referencewrite=reference.aux \fi}
%
{\catcode`\^^M=\active 
  \gdef\obeyendofline{\catcode`\^^M\active \let^^M\ }}%
%
{\catcode`\^^M=\active 
  \gdef\ignoreendofline{\catcode`\^^M=5}}
{\obeyendofline\gdef\rw@start#1{\def\t@st{#1}\ifx\t@st\blankend%
\endgroup {\@sf} \relax \else \ifx\t@st\bl@nkend \endgroup {\@sf} \relax%
\else \rw@begin#1
\backtotext
\fi \fi } }
{\obeyendofline\gdef\rw@begin#1
{\def\n@xt{#1}\rw@toks={#1}\relax%
\rw@next}}
\def\blankend{}
{\obeylines\gdef\bl@nkend{
}}
\newif\iffirstrefline  \firstreflinetrue
\def\rwr@teswitch{\ifx\n@xt\blankend \let\n@xt=\rw@begin %
 \else\iffirstrefline \global\firstreflinefalse%
\immediate\write\rw@write{\noexpand\obeyendofline \the\rw@toks}%
\let\n@xt=\rw@begin%
      \else\ifx\n@xt\rw@@d \def\n@xt{\immediate\write\rw@write{%
	\noexpand\ignoreendofline}\endgroup \@sf}%
	     \else \immediate\write\rw@write{\the\rw@toks}%
	     \let\n@xt=\rw@begin\fi\fi \fi}
\def\rw@next{\rwr@teswitch\n@xt}
\def\rw@@d{\backtotext} \let\rw@end=\relax
\let\backtotext=\relax

\newdimen\refindent	\refindent=20pt
\newmuskip\refskip
\newmuskip\regularrefskip \regularrefskip=2mu
\newmuskip\specialrefskip \specialrefskip=-2mu
\def\refattach#1{\@sf \ifhmode\ifnum\spacefactor=1250 \refskip=\specialrefskip
 \else\ifnum\spacefactor=3000 \refskip=\specialrefskip
 \else\ifnum\spacefactor=1001 \refskip=\specialrefskip
 \else \refskip=\regularrefskip \fi\fi\fi
 \else \refskip=\regularrefskip \fi
 \ref@ttach{\strut^{\mkern\refskip #1}}}
\def\ref@ttach#1{\ifmmode #1\else\ifhmode\unskip${}#1$\relax\fi\fi{\@sf}}
\def\PLrefmark#1{ [#1]{\@sf}}
\def\NPrefmark#1{\refattach{\scriptstyle [ #1 ] }}
\let\PRrefmark=\refattach
\def\refmark{\keepspacefactor\refm@rk}
\def\refm@rk#1{\relax\therefm@rk{#1}}
\def\originalrefs{\let\therefm@rk=\NPrefmark}
\def\PRrefs{\let\therefm@rk=\PRrefmark \let\therefitem=\PRrefitem}
\def\PLrefs{\let\therefm@rk=\PLrefmark \let\therefitem=\PLrefitem}
\def\PRrefitem#1{\refitem{#1.}}
\def\PLrefitem#1{\refitem{[#1]}}
\let\therefitem=\PRrefitem
\def\refitem#1{\par \hangafter=0 \hangindent=\refindent \Textindent{#1}}
\def\REFNUM#1{\eatspace\keepspacefactor\refch@ck \firstreflinetrue%
 \global\advance\referencecount by 1 \xdef#1{\the\referencecount}}
\def\eatspace{\ifhmode\unskip\fi}
\def\refnum#1{\keepspacefactor\refch@ck \firstreflinetrue%
 \global\advance\referencecount by 1 \xdef#1{\the\referencecount}\refend}
\def\REF#1{\REFNUM#1%
 \immediate\write\referencewrite{%
 \noexpand\therefitem{#1}}%
\begingroup\obeyendofline\rw@start}
\def\ref{\refnum\?%
 \immediate\write\referencewrite{\noexpand\therefitem{\?}}%
\begingroup\obeyendofline\rw@start}
\def\Ref#1{\refnum#1%
 \immediate\write\referencewrite{\noexpand\therefitem{#1}}%
\begingroup\obeyendofline\rw@start}
\def\REFS#1{\REFNUM#1\global\lastrefsbegincount=\referencecount
\immediate\write\referencewrite{\noexpand\therefitem{#1}}%
\begingroup\obeyendofline\rw@start}
\def\refend{\refm@rk{\number\referencecount}}
%

%

\newcount\figurecount	  \figurecount=0
\newif\iffigureopen	  \newwrite\figurewrite
\newdimen\digitwidth \setbox0=\hbox{\rm0} \digitwidth=\wd0
\def\zerophant{\kern\digitwidth}
\def\FIGNUM#1{\keepspacefactor\figch@ck \firstreflinetrue%
\global\advance\figurecount by 1 \xdef#1{\the\figurecount}}
\def\figch@ck{\chardef\rw@write=\figurewrite \iffigureopen\else
   \immediate\openout\figurewrite=figures.aux
   \figureopentrue\fi}
\def\figitem#1{\par\indent \hangindent2\parindent \textindent{Fig. #1\ }}
\def\FIGLABEL#1{\ifnum\number#1<10 \def\figlabel{#1.\zerophant}\else%
\def\figlabel{#1.}\fi}
\def\FIG#1{\FIGNUM#1\FIGLABEL#1%
\immediate\write\figurewrite{\noexpand\figitem{\figlabel}}%
\begingroup\obeyendofline\rw@start}
\def\Figname#1{\FIGNUM#1Fig.~#1\FIGLABEL#1%
\immediate\write\figurewrite{\noexpand\figitem{\figlabel}}%
\begingroup\obeyendofline\rw@start}
\def\fig{\FIGNUM\? fig.~\? \FIGLABEL\?
\immediate\write\figurewrite{\noexpand\figitem{\figlabel}}%
\begingroup\obeyendofline\rw@start}
\def\figure{\FIGNUM\? figure~\? \FIGLABEL\?
\immediate\write\figurewrite{\noexpand\figitem{\figlabel}}%
\begingroup\obeyendofline\rw@start}
\def\Fig{\FIGNUM\? Fig.~\? \FIGLABEL\?
\immediate\write\figurewrite{\noexpand\figitem{\figlabel}}%
\begingroup\obeyendofline\rw@start}
\def\Figure{\FIGNUM\? Figure~\? \FIGLABEL\?
\immediate\write\figurewrite{\noexpand\figitem{\figlabel}}%
\begingroup\obeyendofline\rw@start}
\def\figout{\par \penalty-400 \vskip\chapterskip
  \spacecheck\referenceminspace \immediate\closeout\figurewrite
  \figureopenfalse
  \line{\fourteenrm
   \hfil FIGURE CAPTION\ifnum\figurecount=1 \else S \fi\hfil}
  \vskip\headskip
  \input figures.aux
  }
\newcount\tablecount	 \tablecount=0
\newif\iftableopen	 \newwrite\tablewrite
\def\tabch@ck{\chardef\rw@write=\tablewrite \iftableopen\else
   \immediate\openout\tablewrite=tables.aux
   \tableopentrue\fi}
\def\TABNUM#1{\keepspacefactor\tabch@ck \firstreflinetrue%
\global\advance\tablecount by 1 \xdef#1{\the\tablecount}}
\def\tableitem#1{\par\indent \hangindent2\parindent \textindent{Table #1\ }}
\def\TABLE#1{\TABNUM#1\FIGLABEL#1%
\immediate\write\tablewrite{\noexpand\tableitem{\figlabel}}%
\begingroup\obeyendofline\rw@start}
\def\Table{\TABNUM\? Table~\?\FIGLABEL\?%
\immediate\write\tablewrite{\noexpand\tableitem{\figlabel}}%
\begingroup\obeyendofline\rw@start}
\def\tabout{\par \penalty-400 \vskip\chapterskip
  \spacecheck\referenceminspace \immediate\closeout\tablewrite \tableopenfalse
  \line{\fourteenrm\hfil TABLE CAPTION\ifnum\tablecount=1 \else S\fi\hfil}
  \vskip\headskip
  \input tables.aux
  }
\PRrefs
\def\etal{{\it et al.}}
%
\def\masterreset{\global\pagenumber=1 \global\chapternumber=0
   \global\equanumber=0 \global\sectionnumber=0
   \global\referencecount=0 \global\figurecount=0 \global\tablecount=0 }
\def\FRONTPAGE{\ifvoid255\else\vfill\penalty-2000\fi
      \masterreset\global\frontpagetrue
      \global\lastf@@t=0 \global\footsymbolcount=0}

\def\papersize{\hsize=35pc\vsize=50pc\hoffset=1pc\voffset=6pc
  \skip\footins=\bigskipamount}
\def\paperstyle{\normalspace\papersize}
\paperstyle
\newskip\frontpageskip
\newtoks\Pubnum \newtoks\pubnum
\newtoks\s@condpubnum \newtoks\th@rdpubnum
\newif\ifs@cond \s@condfalse
\newif\ifth@rd \th@rdfalse
\newif\ifp@bblock  \p@bblocktrue
\newcount\Year 
\def\Yearset{\Year=\year \advance\Year by -1900 
 \ifnum\month<4 \advance\Year by -1 \fi}
\def\PH@SR@V{\doubl@true \baselineskip=24.1pt plus 0.2pt minus 0.1pt
	     \parskip= 3pt plus 2pt minus 1pt }
\def\PHYSREV{\paperstyle\PhysRevtrue\PH@SR@V}
\def\titlepage{\Yearset\FRONTPAGE\paperstyle\ifPhysRev\PH@SR@V\fi
   \ifp@bblock\p@bblock\fi}
\def\nopubblock{\p@bblockfalse}
\def\endpage{\vfil\break}
\frontpageskip=1\medskipamount plus .5fil
\Pubnum={TU--\the\pubnum }
\pubnum={ }
\def\secondpubnum#1{\s@condtrue\s@condpubnum={#1}}
\def\thirdpubnum#1{\th@rdtrue\th@rdpubnum={#1}}
\def\p@bblock{\begingroup \tabskip=\hsize minus \hsize
   \baselineskip=1.5\ht\strutbox \topspace-2\baselineskip
   \halign to\hsize{\strut ##\hfil\tabskip=0pt\crcr
   \the\Pubnum\cr 
   \ifs@cond \the\s@condpubnum\cr\fi
   \ifth@rd \the\th@rdpubnum\cr\fi 
   \the\Month \cr}\endgroup}
\def\title#1{\hrule height0pt depth0pt
   \vskip\frontpageskip \titlestyle{#1} \vskip\headskip }
\def\author#1{\vskip\frontpageskip\titlestyle{\twelvecp #1}\nobreak}

\def\address#1{\par\kern 5pt\titlestyle{\twelvepoint\it #1}}
\def\andaddress{\par\kern 5pt \centerline{\sl and} \address}
\def\abstract{\vskip\frontpageskip\centerline{\fourteenrm ABSTRACT}
 \vskip\headskip }

%
%
%

\def\\{\relax\ifmmode\backslash\else$\backslash$\fi}
\def\globaleqnumbers{\relax\if\equanumber<0\else\global\equanumber=-1\fi}
\def\nextline{\unskip\nobreak\hskip\parfillskip\break}

\def\cropen#1{\crcr\noalign{\vskip #1}}
\def\crr{\cropen{10pt}}
\def\topspace{\hrule height 0pt depth 0pt \vskip}

\let\int=\intop 
\def\prop{\mathrel{{\mathchoice{\pr@p\scriptstyle}{\pr@p\scriptstyle}%
 {\pr@p\scriptscriptstyle}{\pr@p\scriptscriptstyle} }}}
\def\pr@p#1{\setbox0=\hbox{$\cal #1 \char'103$}
   \hbox{$\cal #1 \char'117$\kern-.4\wd0\box0}}
\def\lsim{\mathrel{\mathpalette\@versim<}}
\def\gsim{\mathrel{\mathpalette\@versim>}}
\def\@versim#1#2{\lower0.2ex\vbox{\baselineskip\z@skip\lineskip\z@skip
  \lineskiplimit\z@\ialign{$\m@th#1\hfil##\hfil$\crcr#2\crcr\sim\crcr}}}
%
%
%
\let\sec@nt=\sec
\def\sec{\relax\ifmmode\let\n@xt=\sec@nt\else\let\n@xt\section\fi\n@xt}
\def\obsolete#1{\message{Macro \string #1 is obsolete.}}
\def\firstsec#1{\obsolete\firstsec \section{#1}}
\def\firstsubsec#1{\obsolete\firstsubsec \subsection{#1}}
\def\thispage#1{\obsolete\thispage \global\pagenumber=#1\frontpagefalse}
\def\thischapter#1{\obsolete\thischapter \global\chapternumber=#1}
\def\nextequation#1{\obsolete\nextequation \global\equanumber=#1
   \ifnum\the\equanumber>0 \global\advance\equanumber by 1 \fi}
\def\BOXITEM{\afterassigment\B@XITEM\setbox0=}
\def\B@XITEM{\par\hangindent\wd0 \noindent\box0 }
%

%
\catcode`\@=12 
\message{Done }
\everyjob{\relax
 }
\relax

\font\fourteenbi=cmmib10 scaled\magstep2   \skewchar\fourteenbi='177
\font\twelvebi=cmmib10 scaled\magstep1     \skewchar\twelvebi='177
\font\elevenbi=cmmib10 scaled\magstephalf  \skewchar\elevenbi='177
\font\tenbi=cmmib10                        \skewchar\tenbi='177
\font\fourteenbsy=cmbsy10 scaled\magstep2  \skewchar\fourteenbsy='60
\font\twelvebsy=cmbsy10 scaled\magstep1    \skewchar\twelvebsy='60
\font\elevenbsy=cmbsy10 scaled\magstephalf \skewchar\elevenbsy='60
\font\tenbsy=cmbsy10                       \skewchar\tenbsy='60
\font\fourteenbsl=cmbxsl10 scaled\magstep2 
\font\twelvebsl=cmbxsl10 scaled\magstep1 
\font\elevenbsl=cmbxsl10 scaled\magstephalf 
\font\tenbsl=cmbxsl10 
\font\fourteenbit=cmbxti10 scaled\magstep2
\font\twelvebit=cmbxti10 scaled\magstep1
\font\elevenbit=cmbxti10 scaled\magstephalf
\font\tenbit=cmbxti10 
\catcode\lq\@=11 
\def\fourteenbold{\relax
    \textfont0=\fourteenbf	    \scriptfont0=\tenbf
    \scriptscriptfont0=\sevenbf
     \def\rm{\fam0 \fourteenbf \f@ntkey=0 }\relax
    \textfont1=\fourteenbi	    \scriptfont1=\tenbi
    \scriptscriptfont1=\seveni
     \def\oldstyle{\fam1 \fourteenbi\f@ntkey=1 }\relax
    \textfont2=\fourteenbsy	    \scriptfont2=\tenbsy
    \scriptscriptfont2=\sevensy
    \textfont3=\fourteenex     \scriptfont3=\fourteenex
    \scriptscriptfont3=\fourteenex
    \def\it{\fam\itfam \fourteenbit\f@ntkey=4 }\textfont\itfam=\fourteenbit
    \def\sl{\fam\slfam \fourteenbsl\f@ntkey=5 }\textfont\slfam=\fourteenbsl
    \scriptfont\slfam=\tensl
    \def\bf{\fam\bffam \fourteenrm\f@ntkey=6 }\textfont\bffam=\fourteenrm
    \scriptfont\bffam=\tenrm	 \scriptscriptfont\bffam=\sevenrm
    \def\tt{\fam\ttfam \twelvett \f@ntkey=7 }\textfont\ttfam=\twelvett
    \h@big=11.9\p@ \h@Big=16.1\p@ \h@bigg=20.3\p@ \h@Bigg=24.5\p@
    \def\caps{\fam\cpfam \twelvecp \f@ntkey=8 }\textfont\cpfam=\twelvecp
    \setbox\strutbox=\hbox{\vrule height 12pt depth 5pt width\z@}\relax
    \samef@nt}
\def\twelvebold{\relax
    \textfont0=\twelvebf	  \scriptfont0=\ninebf
    \scriptscriptfont0=\sevenbf
     \def\rm{\fam0 \twelvebf \f@ntkey=0 }\relax
    \textfont1=\twelvebi	  \scriptfont1=\ninei
    \scriptscriptfont1=\seveni
     \def\oldstyle{\fam1 \twelvebi\f@ntkey=1 }\relax
    \textfont2=\twelvebsy	  \scriptfont2=\ninesy
    \scriptscriptfont2=\sevensy
    \textfont3=\twelveex	  \scriptfont3=\twelveex
    \scriptscriptfont3=\twelveex
    \def\it{\fam\itfam \twelvebit \f@ntkey=4 }\textfont\itfam=\twelvebit
    \def\sl{\fam\slfam \twelvebsl \f@ntkey=5 }\textfont\slfam=\twelvebsl
    \scriptfont\slfam=\ninesl
    \def\bf{\fam\bffam \twelverm \f@ntkey=6 }\textfont\bffam=\twelverm 
    \scriptfont\bffam=\ninerm	  \scriptscriptfont\bffam=\sevenrm
    \def\tt{\fam\ttfam \twelvett \f@ntkey=7 }\textfont\ttfam=\twelvett
    \h@big=10.2\p@ \h@Big=13.8\p@ \h@bigg=17.4\p@ \h@Bigg=21.0\p@
    \def\caps{\fam\cpfam \twelvecp \f@ntkey=8 }\textfont\cpfam=\twelvecp
    \setbox\strutbox=\hbox{\vrule height 10pt depth 4pt width\z@}\relax
    \samef@nt}
\def\elevenbold{\relax
    \textfont0=\elevenbf	  \scriptfont0=\ninebf
    \scriptscriptfont0=\sixbf
     \def\rm{\fam0 \elevenbf \f@ntkey=0 }\relax
    \textfont1=\elevenbi	  \scriptfont1=\ninei
    \scriptscriptfont1=\sixi
     \def\oldstyle{\fam1 \elevenbi\f@ntkey=1 }\relax
    \textfont2=\elevenbsy	  \scriptfont2=\ninesy
    \scriptscriptfont2=\sixsy
    \textfont3=\elevenex	  \scriptfont3=\elevenex
    \scriptscriptfont3=\elevenex
    \def\it{\fam\itfam \elevenbit \f@ntkey=4 }\textfont\itfam=\elevenbit
    \def\sl{\fam\slfam \elevenbsl \f@ntkey=5 }\textfont\slfam=\elevenbsl
    \scriptfont\slfam=\ninesl
    \def\bf{\fam\bffam \elevenrm \f@ntkey=6 }\textfont\bffam=\elevenrm 
    \scriptfont\bffam=\ninerm	  \scriptscriptfont\bffam=\sixrm
    \def\tt{\fam\ttfam \eleventt \f@ntkey=7 }\textfont\ttfam=\eleventt
    \h@big=9.311\p@ \h@Big=12.6\p@ \h@bigg=15.88\p@ \h@Bigg=19.17\p@
    \def\caps{\fam\cpfam \elevencp \f@ntkey=8 }\textfont\cpfam=\elevencp
    \setbox\strutbox=\hbox{\vrule height 9pt depth 4pt width\z@}\relax
    \samef@nt}
\def\tenbold{\relax
    \textfont0=\tenbf	       \scriptfont0=\sevenrm
    \scriptscriptfont0=\fiverm
    \def\rm{\fam0 \tenrm \f@ntkey=0 }\relax
    \textfont1=\tenbi	       \scriptfont1=\seveni
    \scriptscriptfont1=\fivei
    \def\oldstyle{\fam1 \tenbi \f@ntkey=1 }\relax
    \textfont2=\tenbsy	       \scriptfont2=\sevensy
    \scriptscriptfont2=\fivesy
    \textfont3=\tenex	       \scriptfont3=\tenex
    \scriptscriptfont3=\tenex
    \def\it{\fam\itfam \tenbit \f@ntkey=4 }\textfont\itfam=\tenbit
    \def\sl{\fam\slfam \tenbsl \f@ntkey=5 }\textfont\slfam=\tenbsl
    \def\bf{\fam\bffam \tenrm \f@ntkey=6 }\textfont\bffam=\tenrm
    \scriptfont\bffam=\sevenrm  \scriptscriptfont\bffam=\fiverm
    \def\tt{\fam\ttfam \tentt \f@ntkey=7 }\textfont\ttfam=\tentt
    \def\caps{\fam\cpfam \tencp \f@ntkey=8 }\textfont\cpfam=\tencp
    \h@big=8.5\p@ \h@Big=11.5\p@ \h@bigg=14.5\p@ \h@Bigg=17.5\p@ 
    \setbox\strutbox=\hbox{\vrule height 8.5pt depth 3.5pt width\z@}\relax
    \samef@nt}
\def\bold{\iftwelv@\twelvebold\else\ifelev@n\elevenbold\else\tenbold\fi\fi}
\catcode\lq\@=12
\font\seventeenbi=cmmib10 scaled\magstep3 
\def\jnfont{\rm}
\def\AP#1,{{\jnfont Ann.\ Phys.\ (N.Y.)} {\bf #1},}
\def\JETPL#1,{{\jnfont JETP Lett.}\ {\bf #1},}
\def\NPB#1,{{\jnfont Nucl.\ Phys.}\ {\bf B#1},}
\def\PL#1,{{\jnfont Phys.\ Lett.}\ {\bf #1},}
\def\PLB#1,{{\jnfont Phys.\ Lett.\ B}~{\bf #1},}
\def\PRD#1,{{\jnfont Phys.\ Rev.\ D}~{\bf #1},}
\def\PRL#1,{{\jnfont Phys.\ Rev.\ Lett.}\ {\bf #1},}
\def\SJNP#1,{{\jnfont Sov.\ J. Nucl.\ Phys.}\ {\bf #1},}
\def\ZPC#1,{{\jnfont Z. Phys.\ C} {\bf #1},}
\advance\vsize by 36pt
\advance\voffset by -36pt
\PLrefs
\pubnum{497}
\secondpubnum{TIT-HEP-308}
\finaltrue
\let\to=\rightarrow
\mathchardef\REAL="023C
\mathchardef\IMAG="023D
\def\Re{\REAL\!{\it e}} 
 
\def\r{\noalign{\vskip 3\jot }}
\def\Li#1(#2){\mathop{\rm Li}\nolimits_{#1}\left({#2}\right)}
\def\pe{p_e}
\def\pebar{\bar p_e}
\def\zetaorzetabar{^{^(}\!\bar\zeta{}^{^)}}
\newcount\footnotecount \footnotecount=0
\let\ofootnote=\footnote
\def\footnote{\global\advance\footnotecount by 1 
 \ofootnote{\the\footnotecount}}
\def\chapter#1{\par \penalty-300 \vskip\chapterskip
   \spacecheck\chapterminspace
   \chapterreset \par\noindent{\twelvebold\chapterlabel \ #1}\par 
   \nobreak\vskip\headskip \penalty 30000
   \wlog{\string\chapter\ \chapterlabel} }
\chapterminspace=120pt
\def\ack{\par\penalty-100\bigskip \spacecheck\sectionminspace
   \par\noindent{\twelvebold Acknowledgments}\par
   \nobreak\vskip\headskip }
\def\APPENDIX#1#2{\par\penalty-300\vskip\chapterskip
   \spacecheck\chapterminspace \chapterreset \xdef\chapterlabel{#1}
   \par\noindent{\twelvebold Appendix #2}\par 
   \nobreak\vskip\headskip \penalty 30000
   \wlog{\string\Appendix\ \chapterlabel} }
\def\Appendix#1#2{\par\penalty-300\vskip\chapterskip
   \spacecheck\chapterminspace \chapterreset \xdef\chapterlabel{#1}
   \par\noindent{\twelvebold Appendix #1.\ #2}\par 
   \nobreak\vskip\headskip \penalty 30000
   \wlog{\string\Appendix\ \chapterlabel} }

\def\figout{\par \penalty-400 \vskip\chapterskip
   \spacecheck\referenceminspace \immediate\closeout\figurewrite
   \figureopenfalse
   \leftline{\twelvebold Figure Captions}\par
   \nobreak\vskip\headskip \penalty 30000
   \input figures.aux
   }
\titlepage
\title{\begingroup\textfont1=\seventeenbi \scriptfont1=\twelvebi 
\scriptfont0=\twelvebf \scriptfont2=\twelvebsy 
\seventeenbf Hard gluon emission from colored scalar pairs\break 
in $e^+e^-$ annihilation\endgroup}

\author{\fourteenrm Ken-ichi Hikasa}
\address{Department of Physics, Tohoku University\break 
Aoba-ku, Sendai 980-77, Japan}
\author{\fourteenrm Junji Hisano}
\address{Department of Physics, Tokyo Institute of Technology\break
Oh-okayama, Meguro-ku, Tokyo 152, Japan}

\abstract

We study QCD correction to the pair production of colored scalar 
particles in electron-positron annihilation
with an emphasis on gluon emission in the final state.
We discuss the usefulness of working in a ``quasi-two-body'' frame and 
present the helicity amplitudes for the process.  
We compare the final state configuration with fermion pair production 
and find that the three-jet fraction for the scalars shows quantitative 
difference from that for fermions.  

\endpage

\chapter{Introduction}

The known elementary particles are either spin-1/2 fermions 
(quarks and leptons) or spin-1 gauge bosons.  No elementary spin-0 particle 
has been found to date.  Yet, scalar particles appear in most field theory 
models of particles.  Higgs bosons are the key ingredient for 
electroweak symmetry breaking in the standard model and its extensions.  
Supersymmetry predicts a scalar partner for each fermion degree of freedom, 
thus the existence of three generations of squarks and sleptons.  
Grand unified $E_6$ model contains  colored scalar particles which may be 
interpreted either as leptoquarks or diquarks.  

Many of these scalars can be pair produced in $e^+e^-$ annihilation with a 
cross section comparable to or somewhat smaller than that for fermions.  
Experimental searches for these particles have been extensively performed%
\ref{Particle Data Group, L.~Montanet \etal, \PRD50, 1173 (1994).}.  

\REF\Schwinger{J.~Schwinger, {\it Particles, Sources, and Fields}, Vol.~II 
(Addison-Wesley, Mass., 1973), Chapter 5-4.}
\REF\DH{M.~Drees and K.~Hikasa, \PLB252, 127 (1990).}

If the scalar particle is colored (like the squark and leptoquark), 
the cross section is modified by QCD corrections.  
The calculation of the ${\cal O}(\alpha_s)$ correction is essentially 
the same as scalar QED one-loop calculation which is known for a long time.%
  \footnote{A detailed description can be found in Ref.~\Schwinger.  A 
  typographical error is corrected in Ref.~\DH.}
The ${\cal O}(\alpha_s)$ correction for the scalar pair production is 
numerically quite important.  
In the high energy limit, the total correction 
factor for scalar pair production is {\it four\/} times larger than 
that for fermion pair production\refmark\DH:
$$ \sigma(e^+e^-\to \zeta\bar\zeta(g)) = 
\sigma_0 \biggl( 1 + 3\,{C_{\!R}\alpha_s\over\pi} \biggr) \;, \Eq$$
where $\sigma_0$ is the lowest-order cross section and 
$C_{\!R}$ is the second-order SU(3) Casimir eigenvalue 
(4/3 for the fundamental representation like squarks or leptoquarks).  
We use the notation $\zeta$ to represent a generic colored scalar 
particle in the representation $R$.  
At lower energies, especially in the threshold region, the correction 
factor becomes even larger.

The total ${\cal O}(\alpha_s)$ correction consists of two parts, 
one-loop virtual correction to the lowest order process 
$e^+e^-\to \zeta\bar\zeta$ and real gluon emission correction 
$e^+e^-\to \zeta\bar\zeta g$, both of which are infrared divergent.  
This divergence is cancelled when the two contributions are added.  
Essential for this cancellation is only the part of the three-body phase 
space where the gluon is soft 
(or collinear to one of the scalars when the scalar mass goes to zero).  
The rest of the three-body phase space corresponds to the final state 
with a hard gluon which may be observed as a separate jet 
(``three-jet'' final state).  

The gluon emission process in scalar top quark pair production 
has been studied by Beenakker, H\"opker, and Zerwas\ref{W.~Beenakker, 
R.~H\"opker, and P.~M. Zerwas, \PLB349, 463 (1995).}, who calculated the 
fully differential cross section for $e^+e^-\to \tilde t \bar{\tilde t} g$.   
In the present paper, we derive the cross section in a different Lorentz 
frame (``quasi-two-body'' frame) in which the helicity amplitudes have a 
simple interpretation.  We then extend their analysis and compare the 
three-jet cross sections with those for scalar and fermion pair production 
processes.  

The QCD correction we calculate in this paper is just the effects arising 
from gluon gauge interactions.  
In the supersymmetric standard model there are additional interactions with 
the same strength involving gluinos.  The gluino coupling contribution 
to the virtual ${\cal O}(\alpha_s)$ correction is discussed by 
Arhrib, Capdequi-Peyranere, and Djouadi%
\ref{A.~Arhrib, M.~Capdequi-Peyranere, and A.~Djouadi, \PRD52, 1404 (1995).}.

\chapter{Lowest-Order Cross Section}

We consider the process $e^+e^-\to \zeta\bar\zeta$ which occurs via 
$s$-channel $\gamma$ or $Z$ exchange.   Additional $t$ or $u$-channel 
exchanges have to be included for selectron (or electron sneutrino) 
production or a leptoquark if it couples to electrons with a nonnegligible 
Yukawa coupling.  

In this section we list the lowest-order amplitude and cross section 
for completeness.  The Feynman graph is shown in Fig.~1.  
The amplitude is found to be (see Fig.~1 for the momentum assignments)
%
%
$$ {\cal M}_0  = {e^2\over s} \,H^\mu\,(p-\bar p)_\mu \;,\Eq$$ 
with
$$ H^\mu =  -Q_\zeta\, \bar v(\pebar) \gamma^\mu  u(\pe) 
+ {T_{3\zeta}-Q_\zeta\sin^2\!\theta_W\over\cos^2\!\theta_W\sin^2\!\theta_W} \,
{s\over s-m_Z^2} \,\bar v(\pebar) \gamma^\mu (v_e - a_e\gamma_5) u(\pe) 
\;.\Eq$$
Here $s$ is the $e^+e^-$ c.m.\ energy squared, 
$Q_\zeta$ and $T_{3\zeta}$ are the electric charge and the third component 
of weak isospin of $\zeta$, and 
$$ v_e = -{1\over4} + \sin^2\!\theta_W\;,\qquad a_e = -{1\over4} \;.\Eq$$
We will neglect the electron mass throughout the paper.  

For longitudinally polarized electrons with polarization $P$ colliding 
with unpolarized positrons, the cross section can be written   
$$d\sigma = {1+P\over 4}\, d\sigma_+
+ {1-P\over 4}\, d\sigma_- \;,\Eq$$
%
%
where $d\sigma_\pm$ denotes the cross section for the $e^-_R e^+_L$ and 
$e^-_L e^+_R$ initial states.  We find 
$$ {d\sigma_\pm\over d\cos\theta} = {\pi d_R\alpha^2\beta^3\over 2s} H_\pm^2 
\,\sin^2\!\theta \;,\Eq$$
with
$$ H_\pm =  -Q_\zeta 
+ {(v_e {\mp} a_e)(T_{3\zeta} {-} Q_\zeta\sin^2\!\theta_W) 
\over \cos^2\!\theta_W \sin^2\!\theta_W}{s\over s-m_Z^2} \;.\Eq$$
%
%
Here,  $d_R$ is the dimension of the color SU(3) representation of $\zeta$ 
(3 for the fundamental representation), and 
$$\beta = \sqrt{1-4m^2/s} \;,\Eq$$
with $m$ being the $\zeta$ mass.  The $\beta^3$ factor reflects the 
$P$ wave threshold.  The total cross section is
$$\sigma_\pm  = {2\pi d_R\alpha^2\beta^3\over 3s} H_\pm^2 \;,\Eq$$
which we will denote by $\sigma_{0\pm}$.

\chapter{${\cal O}(\alpha_s)$ correction}

\section{Virtual one-loop correction}

Feynman diagrams for the process $e^+e^-\to\zeta\bar\zeta$ at 
${\cal O}(\alpha_s)$ are depicted in Fig.~2.  The diagrams (a)--(c) 
are the one-particle-irreducible vertex corrections.  The mixed four-point 
vertex in (b) and (c) 
appears because the scalar kinetic term has two covariant derivatives.  
The diagrams (d) and (e) are $\zeta$ self-energy corrections, and the 
diagrams (f)--(h) show the counterterm contribution.  

We regularize the ultraviolet divergence by dimensional reduction%
  \footnote{Naive dimensional regularization gives the same physical 
   results, although the finite parts of the renormalization constants  
   differ.} 
with $D=4-2\epsilon$, and the infrared singularity by an infinitesimal 
gluon mass $\lambda$.  This procedure does not cause problems with gauge 
invariance since the diagrams do not contain non-Abelian gauge vertices.

We adopt the on-shell renormalization scheme to 
determine the counterterms.  The mass and wave function renormalization 
constants for $\zeta$ are found to be  
$$\eqalignno{
&\displaystyle{\delta m^2\over m^2} = - {C_{\!R}\alpha_s\over 4\pi}
\left[ 3 \left(\displaystyle{1\over\epsilon}  -\gamma_E + \log4\pi \right) 
-3\log \displaystyle{m^2\over\mu^2} + 7 \right] \;,&\Eq\cr
&Z_\zeta-1 = {C_{\!R}\alpha_s\over 4\pi}
\left[ 2 \left(\displaystyle{1\over\epsilon}  -\gamma_E + \log4\pi \right)
-2\log \displaystyle{\lambda^2\over\mu^2} \right] \;,&\Eq\cr}$$
where $\mu$ is the renormalization scale and $\gamma_E$ is the Euler constant. 

We write the $\zeta\bar\zeta\gamma$ vertex function in the form 
$-ieQ_\zeta \Lambda_\mu$ with 
$$ \Lambda_\mu = F(q^2) (p-\bar p)_\mu \;.\Eq$$
(See Fig.~2 for the definition of the momenta.)
Up to the order we work, the $\zeta\bar\zeta Z$ vertex can be 
writtten as 
$${-ie(T_{3\zeta}-Q_\zeta\sin^2\!\theta_W)\over\cos\theta_W\sin\theta_W}
\Lambda_\mu \Eq$$
with the same $\Lambda_\mu$.   
At the lowest order we have  $F(q^2)=1$.  

The counterterm for the vertex can be found using the Ward 
identity.  The same result is obtained by requiring no 
${\cal O}(\alpha_s)$ correction at $q^2=0$, {\it i.e.}, $F(0)=1$.  
Expanding the form factor as 
$$ F(q^2) =  1 + {C_{\!R}\alpha_s\over 2\pi} f(q^2) \;,\Eq$$
we find for $q^2>4m^2$ 
\begingroup
\def\+{{+}}
\def\-{{-}}
$$\eqalignno{
f(q^2) &= 
\biggl(-{1\+\beta^2\over 2\beta} 
\log{1\+\beta\over 1\-\beta} + 1 \biggr) \log{m^2\over\lambda^2} \cr\r
&\quad + {1\+\beta^2\over\beta} \biggl[ \Li2(1\-\beta\over 1\+\beta) 
- \log{2\beta\over 1\+\beta}\log{1\+\beta\over 1\-\beta} 
- {1\over4}\log^2{1\+\beta\over1\-\beta} + {\pi^2\over3} \biggr]\cr\r
&\quad + {1\+\beta^2\over\beta}\log{1\+\beta\over 1\-\beta} -2 \cr\r
&\quad + i\pi \, {1\+\beta^2\over 2\beta} \biggl(\log{m^2\over\lambda^2} 
+ \log{4\beta^2\over 1\-\beta^2} -2 \biggr) \;.&\Eq\cr}$$ 
\REF\tHV{G.~'t~Hooft and M.~Veltman, \NPB153, 365 (1979).}
Here $\mathop{\rm Li_2}(x)$ is the dilogarithm (Spence) function%
\footnote{For a convenient expansion for numerical evaluation of  
dilogarithm, see Ref.~\tHV.} 
$\mathop{\rm Li_2}(x)=-\int_0^x dt\,\log(1-t)/t$.  
At ${\cal O}(\alpha_s)$, the lowest order cross section is multiplied by the 
factor
$$ 1 + {C_{\!R}\alpha_s\over\pi}\, \Re f(s) \;.\Eq$$
The infrared singularity is cancelled by real gluon emission, 
to which we now turn.
\endgroup

\section{Real gluon emission}

The Feynman diagrams for the process $e^+e^-\to \zeta\bar\zeta g$ 
at the lowest order are shown in Fig.~3.  The amplitudes are found to 
be
%
%
$${\cal M} = 
{e^2 g_s T^a\over s}\, H^\mu 
\left[ {(p-\bar p+k)_\mu p_\alpha \over  p{\cdot}k}
- {(p-\bar p-k)_\mu \bar p_\alpha \over \bar p{\cdot}k}
- 2g_{\mu\alpha} \right] \;\epsilon^{*\alpha} \;.\Eq$$
Here $T^a$ is the SU(3) generator in the representation $R$.

We calculate the cross section for this process in two methods, with 
agreeing results.  One is the conventional trace technique.  The other 
method is a helicity amplitude technique described in Appendix A.  

The differential cross section may be parametrized by four variables.  
We find it convenient to use the following four:

\itemitem{$\tau$ :} $\zeta\bar\zeta$ c.m.\ energy squared normalized to $s$, 
$\tau=(p+\bar p)^2/s$;
\itemitem{$\Theta$ :} angle between the $e^-$ and the gluon 
in the $e^+e^-$ c.m.\ frame;
\itemitem{$\theta$ :} angle between the $\zeta$ and the gluon 
in the $\zeta\bar\zeta$ c.m.\ frame;
\itemitem{$\phi$ :} azimuthal angle between the $e^-$ and the $\zeta$ with 
respect to the gluon, measured from the $e^-$ to the $\zeta$ direction 
(common to both frames).  

The use of these variables is motivated by the fact that the amplitude 
can be split into two subprocesses $e^+e^-\to V^*$ ($V^*=\gamma$, $Z$) 
and $V^*\to \zeta\bar\zeta g$, and the latter process can be most 
conveniently evaluated in the $\zeta\bar\zeta$ c.m.\ frame.  The detail 
and the result for the helicity amplitudes may be found in Appendix A.  

The polarized cross section is 
$$\eqalignno{
&{d\sigma_\pm \over d\tau\,d\cos\Theta\,d\cos\theta\,d\phi} 
=  {d_R\alpha^2 H_\pm^2 \over 8 s}\,{C_{\!R}\alpha_s\over\pi}\, v (1-\tau)\cr\r
&\qquad\times\biggl\lbrace 
A (1+\cos^2\Theta) + B (1-3\cos^2\Theta )
+ C \sin\Theta\cos\Theta\cos\phi + D \sin^2\Theta \cos2\phi 
\biggr\rbrace\;,\cr&&\Eq\cr}$$
where
$$\eqalignno{
A &= 1 + {2\beta^2 \tau v^2 \sin^2\!\theta\over 
(1-\tau)^2 (1-v^2\cos^2\!\theta)^2} \;,&\Eqa\cr\r
B &= {2 \tau v^4 \sin^2\!\theta \cos^2\!\theta \over 
(1-\tau)^2 (1-v^2\cos^2\!\theta)^2} \;,&\Eqb\cr\r
C &= -{4\sqrt\tau v^2\sin\theta\cos\theta 
\bigl[ \beta^2-v^2+(1+\tau)v^2\sin^2\!\theta\bigr] 
\over (1-\tau)^2(1-v^2\cos^2\!\theta)^2} \;,&\Eqc\cr\r
D &= -{2\tau v^2\sin^2\!\theta(\beta^2-v^2\cos^2\!\theta)\over 
(1-\tau)^2(1-v^2\cos^2\!\theta)^2}  \;.&\Eqd\cr}$$
Here 
$$ v = \sqrt{ 1 - 4m^2/ \tau s} \Eq$$
is the velocity of $\zeta$ in the $\zeta\bar\zeta$ c.m.\ frame.  

If one integrates the cross section over the whole 3-body phase space, 
one encounters infrared divergence coming from soft-gluon region.  We 
separate this region from the rest by the condition that the gluon 
energy $k^0<\omega$, where the cutoff $\omega$ is infinitesimally small.  

In this region, we regularize the divergence by giving an infinitesimal mass 
$\lambda$ to the gluon, which satisfies $\lambda\ll\omega$.  We recalculate 
the amplitude with finite mass.  The amplitude factorizes into the 
lowest-order part and the soft gluon factor.  
Since a very soft gluon does not alter the kinematics of the $\zeta\bar\zeta$ 
state, we can also integrate 
over the soft gluon phase space ignoring momentum conservation.  
Integrate over this soft-gluon region, we find the cross section
\begingroup
\def\+{{+}}
\def\-{{-}}
$$\eqalignno{\sigma_{\rm soft} &= \sigma_0 {C_{\!R}\alpha_s\over\pi}\,
\Biggl\{ \biggl({1\+\beta^2\over2\beta} \log{1\+\beta\over 1\-\beta} -1\biggr)
          \log{4\omega^2\over\lambda^2}
\crr &\qquad\quad
         +{1\+\beta^2\over\beta}\biggl[
           \Li2({1\-\beta\over 1\+\beta}) 
         - \log{2\beta\over1\+\beta}\log{1\+\beta\over 1\-\beta} 
           -{1\over4}\log^2{1\+\beta\over 1\-\beta} - {\pi^2\over6} \biggl]
\crr &\qquad\quad
        + {1\over\beta}\log{1\+\beta\over 1\-\beta} 
         \Biggr\} \;,&\Eq\cr}$$
where $\sigma_0$ is the lowest-order cross section.

For the rest of the phase space, we can set the gluon mass to zero.  
We first integrate over $\Theta$, $\phi$ to find
$${d\sigma_\pm\over d\tau\,d\cos\theta} 
= \sigma_{0\pm}\,{C_{\!R}\alpha_s\over\pi }\,{v(1-\tau)\over \beta^3}\,
\biggl[1 + {2\beta^2 \tau v^2 \sin^2\!\theta\over 
(1-\tau)^2 (1-v^2\cos^2\!\theta)^2}\biggr] \;,\Eq$$
which we will use to study the event configuration in the next section.  
The integrated cross section is 
$$\eqalignno{\sigma_{\rm hard} &= \sigma_0 {C_{\!R}\alpha_s\over\pi}\,
\Biggl\{ \biggl({1+\beta^2\over2\beta} \log{1\+\beta\over 1\-\beta} -1\biggr)
          \log{m^2\over4\omega^2}
\crr &\qquad\quad
         +{1+\beta^2\over\beta}\biggl[
          2\,\Li2({1\-\beta\over 1\+\beta})
         + 2\,\Li2(-{1\-\beta\over 1\+\beta})
          -\log{2\over1\+\beta}\log{1\+\beta\over 1\-\beta}
\crr &\qquad\qquad\qquad\quad
          + {1\over2} \log^2{1\+\beta\over 1\-\beta} - {\pi^2\over6} \biggr]
 -3\log{4\over 1\-\beta^2} - 4\log\beta
\crr &\qquad\quad
        - {1\over4\beta^3}\, (3+\beta^2)(1-\beta^2) 
          \log{1\+\beta\over 1\-\beta} 
        + {1\over2\beta^2}(3+7\beta^2) 
         \Biggr\} \;.&\Eq\cr}$$

The $\omega$ dependence cancels out when these two cross sections are summed.  
The infrared divergence is cancelled by adding the virtual one-loop 
correction discussed in the previous subsection.

\section{Total correction}

The ${\cal O}(\alpha_s)$ corrected cross section for 
$e^+e^-\to \zeta\bar\zeta (g)$ can thus be written as
$$\sigma=\sigma_0\,\biggl[ 1 + {C_{\!R}\alpha_s\over\pi}\Delta \biggr] \;,\Eq$$
where
$$ \Delta = {1\over\beta}\,A(\beta)
       + {1\over4\beta^3}(-3\+10\beta^2\+5\beta^4)\log{1\+\beta\over 1\-\beta}
       + {3\over2\beta^2}(1\+\beta^2)  \;,\Eq$$
with
$$\eqalignno{
A(\beta) &= (1+\beta^2)\biggl[
          4 \,\Li2({1\-\beta\over 1\+\beta}) 
        + 2 \,\Li2(-{1\-\beta\over 1\+\beta})
        - 3\log{2\over 1\+\beta}\log{1\+\beta\over 1\-\beta}
\crr &\qquad\qquad\quad
       - 2 \log\beta\log{1\+\beta\over 1\-\beta} \biggr]
 - 3\beta\log{4\over 1 \- \beta^2}  - 4\beta \log\beta\;.&\Eqn\eqA\cr}$$
This may be compared with the corresponding formulas for the pair production 
of a colored fermion pair via vector and axial vector currents (the detail 
on the latter may be found in Appendix C)
$$\eqalignno{ \Delta_V &= {1\over\beta}\,A(\beta)
       + {33+22\beta^2-7\beta^4\over 8\beta(3-\beta^2)}
            \log{1\+\beta\over 1\-\beta}
       + {3(5-3\beta^2)\over 4(3-\beta^2)}  \;.&\Eq\cr\r
\Delta_A &= {1\over\beta}\,A(\beta)
       + {1\over 32\beta^3}\,(21+59\beta^2+19\beta^4-3\beta^6)
            \log{1\+\beta\over 1\-\beta} \cr\r
&\quad + {3\over 16\beta^2}\,(-7+10\beta^2+\beta^4)  \;.&\Eq\cr}$$

The dependence on $\beta$ of the cross section normalized to its 
lowest order value is shown in Fig.~4.  The corresponding quantities  
for fermion pair production are also shown for comparison.  
The correction for scalars is larger than that for fermions via vector 
current for all values of $\beta$.  The correction for fermions via 
axial current is equal to that for vector current at the high energy 
limit (which reflects the chiral symmetry in this limit), but approaches 
the scalar curve at small $\beta$.  This latter behavior reflects the fact 
that spin does not play an important role in the nonrelativistic region.   
The difference of the vector current result is due to its $S$-wave 
dynamics contrast to the $P$ wave behavior of the others.  

At the high energy limit $\beta\to 1$, one finds
$$ \Delta = 3 \;,\qquad \Delta_V=\Delta_A={3\over4} \;.\Eq$$
The correction for the scalar pair is four times larger.  
Near the threshold, one has 
$$ \Delta \simeq \Delta_A \simeq {\pi^2\over2\beta}-2 \;,\qquad 
   \Delta_V \simeq {\pi^2\over2\beta}-4 \;.\Eq$$
The terms proportional to $1/\beta$ are the consequence of Coulomb gluon 
exchange.  

It is well known that perturbation expansion in $\alpha_s$ 
breaks down very near the threshold and one has to sum over ladder 
Coulomb gluon diagrams\refmark{\Schwinger}.  
For the case of scalar pair production with the $P$ wave threshold 
behavior, the lowest-order cross section is propotional to $\beta^3$, 
which is modified to $\beta^2$ by the ${\cal O}(\alpha_s)$ correction.  
The nonperturbative contribution further enhances the cross section 
and leads to a constant cross section at the threshold\Ref\BFK{I.~I. Bigi, 
V.~S. Fadin, and V.~Khoze, \NPB377, 461 (1992).}.  
In the nonrelativistic approximation with the Coulomb potential (one gluon 
exchange approximation), the cross section is
$$ \sigma_\pm = {2\pi^2 d_R\alpha^2\over 3s}\,H_\pm^2 \,
{C_{\!R}\alpha_s\bigl(\beta^2+{1\over4}C_{\!R}^2\alpha_s^2\bigr)
\over \bigl[1 -\exp(-\pi C_{\!R}\alpha_s/\beta)\bigr]} \;,\Eq$$
%
%
which gives at the threshold 
$$ \Delta = {\pi^2 C_{\!R}^2\alpha_s^2\over 4 \beta^3} \;.\Eq$$
This effect becomes important only at very near threshold, $\beta\sim 0.1$, 
however.

\endgroup

If one approaches closer to the threshold, one will encounter $P$-wave 
$\zeta\bar\zeta$ bound states, similar to charmonium and bottomonium 
resonances.  
These states are expected to be less pronounced than the fermion bound 
states, since the resonance formation cross section is proportional 
to the derivative of the wave function at the origin and is suppressed 
by a factor of $\alpha_s^2$ compared to the fermion bound states.  
In the Coulomb potential approximation, this subthreshold cross section 
is given by\refmark{\BFK}
$$ \sigma_\pm = {\pi^2 d_R \alpha^2\over 12}\,H_\pm^2 
\bigl(C_{\!R}\alpha_s\bigr)^5 
\mathop{\textstyle\sum}\limits_{n=2}^\infty {n^2-1\over n^5} 
\delta(s-M_n^2) \;,\Eq$$
%
%
where $M_n = 2m - C_{\!R}^2\alpha_s^2 m/4n^2$.  This Coulombic approximation 
ignores the effect of confinement and is not adequate especially for 
higher bound states.  One needs to solve Schr\"odinger equation with 
a realistic color-force potential to obtain the cross section in the 
resonance region.

If $\zeta$ has a short lifetime with $\Gamma \gsim \alpha_s^2 m$, 
it does not live long enough to form a bound state.  
There will be only a smooth bump or shoulder instead of the resonances 
at the threshold.  The situation will be similar to the case for the 
top quark pair production\ref{V.S. Fadin and V.A. Khoze, 
\JETPL46, 525 (1987); \SJNP48, 309 (1988);\nextline
M. Strassler and M. Peskin, \PRD43, 1500 (1991).} 
and the cross section can be calculated quite reliably within perturbative 
QCD.

\chapter{Jet configuration}

In this section we study the shape of the three-body final state 
$\zeta\bar\zeta g$ when the gluon is `visible'.  We compare various 
distributions with those for the more familiar final state of a fermion-pair 
plus a gluon.

To define the final state configuration, it is convenient to use 
Lorentz-invariant kinematic variables.  We use the following 
Dalitz variables
$$ x = {2p{\cdot}k \over s}\;,\qquad 
\bar x = {2\bar p{\cdot}k \over s}\;,\Eq$$
for which the phase space density is constant.  The $\zeta\bar\zeta g$ 
cross section is (we drop the subscript $\pm$ from here on)
$${d\sigma\over dxd\bar x} = \sigma_0\,{C_{\!R}\alpha_s\over\pi}\,{1\over\beta}\,
\left[ {2\over\beta^2} 
- \left({1\over x} + {1\over \bar x}\right) 
+ {1+\beta^2\over 2}\,{1\over x\bar x} 
-{1-\beta^2\over 4} \,\left({1\over x^2}+{1\over \bar x^2}\right)\right]
\;.\Eq$$
The infrared singularity arises when the gluon energy is very small, 
$x\sim\bar x\sim0$.  At the high energy limit, additional singularity 
arises from the region of the phase space in which the gluon momentum 
is parallel to that of $\zeta$ or $\bar\zeta$.  This collinear region 
is characterized as $x\sim 0$ or $\bar x\sim 0$.  In these singular regions, 
it is practically impossible to observe the existence of the extra gluon 
in the final hadron system.  We therefore treat the $\zetaorzetabar+\rm gluon$ 
system as a $\zetaorzetabar$ jet when the $\zetaorzetabar g$ invariant mass 
is sufficiently close to the $\zeta$ mass.

We thus divide the three-body phase space into the two-jet 
and three-jet regions.  The two-jet region is defined as
$$ x < x_c \qquad\hbox{or}\qquad \bar x < x_c \Eq$$
and the rest is the three-jet region.  The two-jet cross section 
is the sum of the $\zeta\bar\zeta$ cross section and the $\zeta\bar\zeta g$ 
cross section in the two-jet region.  

In the high energy limit $\beta\to 1$, the integrated three-jet cross section 
is found to be
$$\sigma_{\rm 3j}/\sigma_0 = 
{C_{\!R}\alpha_s\over\pi}\biggl[ 2 \Li2(x_c) + \log^2 x_c -{\pi^2\over6}
- 2 (1-x_c)\log{1-x_c\over x_c} + (3-2x_c)(1-2x_c) \biggr]\;,\Eq$$
whereas for fermions (both vector and axial) we have
$$\eqalignno{\sigma_{\rm 3j}/\sigma_0 &= 
{C_{\!R}\alpha_s\over\pi}\biggl[ 2 \Li2(x_c) + \log^2 x_c -{\pi^2\over6} \cr\r
&\qquad\qquad 
- {1\over2} (1-x_c)(1-3x_c)\log{1-x_c\over x_c} + {5\over4}(1-2x_c) \biggr]
\;.&\Eq\cr}$$
In Fig.~5, we show the three-jet fraction as a function of $x_c$. 
(Here and in the following figures, we plot $\sigma_{\rm 3j}$ normalized to 
the ${\cal O}(\alpha_s)$ total cross section, not to the lowest-order 
cross section.) 
The three-jet fraction for the scalar is larger than that for the fermion 
at large $x_c$ but becomes smaller at small $x_c\lsim 0.1$.  
The approximate form for small $x_c$ is
$$\sigma_{\rm 3j}/\sigma_0 \simeq {C_{\!R}\alpha_s\over\pi}
\biggl[\log^2 x_c + 2 \log x_c + 3 -{\pi^2\over6}\biggr]
\Eq$$
for scalars, and
$$\sigma_{\rm 3j}/\sigma_0 \simeq {C_{\!R}\alpha_s\over\pi}
\biggl[ \log^2 x_c + {3\over2} \log x_c + {5\over4} -{\pi^2\over6}\biggr]\Eq$$
for fermions.  These predictions are not reliable at very small $x_c$ 
since higher order contributions become important.

The three-jet fraction for finite $m$ has a very complicated expression 
and we do not write down the analytic expression here.  Instead, we show 
the numerical result for $\sigma_{\rm 3j}/\sigma$ for several values of 
$\beta$ in Fig.~6. 
For a fixed value of $x_c$, the three-jet fraction rapidly decreases 
as $\beta$ becomes smaller.  Hard gluon emission is much suppressed 
when one approaches to the threshold.  This suppression comes from the 
available phase space and from the structure of the gluon-emission vertex.  
The large QCD enhancement near the threshold can be attributed to two-jet 
configuration and one can assume that most events in the threshold region 
have no extra jets in experimental searches for new colored scalar particles. 

Toward the threshold region, the three-jet fraction in Fig.~6 for scalar 
pairs becomes similar to that for fermion pairs from axial current.  
The fermion pair from vector current shows a different behavior.  This again 
reflects the fact that spin of the particle is not important in the 
nonrelativistic region.  

In jet analyses in $e^+e^-$ physics, one normally uses a jet-defining 
algorithm by combining momentum of hadrons and compares the result with 
the corresponding quantity in the quark-gluon system.  A commonly adopted 
prescription uses the invariant mass of the hadron system to define a jet.  
To conform with this algorithm, we may use the invariant mass variables
$$ y = {(p+k)^2\over s} \;,\qquad \bar y = {(\bar p+k)^2\over s} \;,\Eq$$
instead of $x$ and $\bar x$.  These variables are related by
$$ y = x +{m^2\over s} \;,\qquad \bar y = \bar x +{m^2\over s}
\;.\Eq$$
The definition of three-jet events in terms of $y$ is
$$ y > y_c \qquad\hbox{and}\qquad \bar y > y_c \;.\Eq$$
The three-jet fraction as a function of $y_c$ is shown in Fig.~7.

So far we have assumed that we can somehow distinguish 
a $\zeta$-jet and a gluon jet.  If this is not the case, we need to 
treat the three final particles in a democratic way and define the 
three-jet region as
$$ y > y_c \qquad\hbox{and}\qquad \bar y > y_c \qquad\hbox{and}\qquad 
y_g > y_c \;,\Eq$$
where $y_g = (p+\bar p)/s$.  Since the region $y_g \to 0$ (or 
$y_g \to 4m^2/s$) contains no singularity, this definition does not 
lead to significant modification of the result for small $y_c$.  
In the high energy limit $\beta\to 1$, the three-jet fraction is found to be
$$\eqalignno{\sigma_{\rm 3j}/\sigma_0 &= 
{C_{\!R}\alpha_s\over\pi}\biggl[ 2 \Li2({y_c\over 1-y_c}) 
+ \log^2{1-y_c\over  y_c} -{\pi^2\over6} \cr\r
&\qquad\qquad 
- 2 (1-2y_c)\log{1-2y_c\over y_c} + 3(1-y_c)(1-3y_c) \biggr] \;,&\Eq\cr}$$
whereas for fermions (both vector and axial) we have
$$\eqalignno{\sigma_{\rm 3j}/\sigma_0 &= 
{C_{\!R}\alpha_s\over\pi}\biggl[  2 \Li2({y_c\over 1-y_c}) 
+ \log^2{1-y_c\over  y_c} -{\pi^2\over6} \cr\r
&\qquad\qquad
- {3\over2} (1-2y_c)\log{1-2y_c\over y_c} + {1\over4}(1-3y_c)(5+3y_c) 
\biggr]\;.&\Eq\cr}$$

\chapter{Conclusions}

We have studied ${\cal O}(\alpha_s)$ QCD correction to colored scalar 
pair production in $e^+e^-$ annihilation.  Emphasis is put on gluon 
emission in the final state.  We compared the configuration of the 
three-jet final state with those in fermion pair production and found 
that the two cross sections are quantitatively quite different.  This 
has to be taken into account in the search and study of production of 
scalar particles such as squarks and leptoquarks at LEP-2 and 
future linear colliders.   

In calculating the helicity amplitudes $e^+e^-\to \zeta\bar\zeta g$, 
we found that it is convenient to decompose the amplitude into two 
subprocesses $e^+e^-\to V^*$ ($V^*=\gamma$, $Z$) and 
$V^*\to \zeta\bar\zeta g$.  The amplitudes take a simple form if one 
works in two different Lorentz frames, the $e^+e^-$ c.m.\ frame for the 
former and the $\zeta\bar\zeta$ c.m.\ frame in the latter.  In particular, 
working in the ``quasi-two-body'' frame in the latter $1\to 3$ subprocess 
allows one to take advantage of the strong constraints of rotational 
invariance on two-body states.

\ack

One of us (J.H.) is a fellow of the Japan Society for the Promotion of 
Science. 
This work is partly supported by Grant-in-Aid for Scientific Research 
(no.~06640369) from the Japan Ministry of Education, Science, Sports, and 
Culture.  

\endpage

\Appendix{A}{Helicity amplitudes for $e^+e^-\to \zeta\bar\zeta g$}

A number of formulations%
\ref{See {\it e.g.}, 
P. de Causmaecker \etal, \NPB206, 53 (1982);\nextline
Zh. Xu, D.-H. Zhang, and L. Chang, \NPB291, 392 (1987);\nextline 

S.~J. Parke and T.~R. Taylor, \PL157B, 81 (1985); 
{\bf174}, 465(E) (1986);\nextline 
A.~Kersch and F.~Scheck, \NPB263, 475 (1986);\nextline 
J.~F. Gunion and Z.~Kunszt, \PL161B, 333 (1985);\nextline 
R.~Kleiss and W.~J. Stirling, \NPB262, 235 (1985);\nextline 
K.~Hagiwara and D.~Zeppenfeld, \NPB274, 1 (1986);\nextline 
F.~A. Berends and W.~Giele, \NPB294, 700 (1987);\nextline 
V.~Barger, J.~Ohnemus, and R.J.N. Phillips, \PRD35, 166 (1987).} 
to calculate helicity amplitudes%
\ref{M.~Jacob and G.~C. Wick, \AP7, 404 (1959).} are found in the literature.  
Most of them are particularly suited for many-body final states and 
numerical evaluation of the amplitudes.  The technique we employ here is 
a method%
\ref{K.~Hikasa, to appear.} 
based on spherical-vector basis, which is suited for manual analytic 
calculations and gives amplitudes with clear physical interpretation.    
The method is most useful for $2\to 2$ body processes (also $1\to 2$, 
$2\to 1$).   In these processes, angular momentum conservation strongly 
constrains the form of the amplitudes.  In our technique, this constraint 
can be explicitly extracted in the form of a Wigner $D$ function which 
depends on the scattering angles.  This part represents 
the variation of the amplitude which is imposed by kinematics.  
The rest of the amplitude contains purely dynamical information.  

In general, the helicity amplitude for the process 
$$ a (p_a, \lambda_a) + b (p_b, \lambda_b) \to 
c (p_c, \lambda_c) + d (p_d, \lambda_d) $$
($\lambda_a$ is the helicity of the particle $a$, {\it etc.}) 
in the c.m.\ frame can be written as 
$$ {\cal M} = \widetilde{\cal M}(E_{\rm c.m.},\cos\theta; \{\lambda\}) \,
d^{J_0}_{\lambda_i,\lambda_f}(\theta) \, e^{i(\lambda_i-\lambda_f)}\phi 
\;, \Eq$$
where $\lambda_i=\lambda_a-\lambda_b$, $\lambda_f=\lambda_c-\lambda_d$, 
$J_0=\max(\lambda_i,\lambda_f)$, and ($\theta$, $\phi$) are the scattering 
angles (in the frame $\vec p_a$ is along the $z$ axis, $\vec p_c$ is 
in the direction ($\theta$, $\phi$)).  The $d^J_{\lambda\lambda'}$ is 
the Wigner $d$ function.  The last two factors in this formula 
represent the ``minimal'' angular distribution imposed by the kinematics.  
The physical meaning of $J_0$ is the smallest allowed angular momentum 
for the process.  The $\cos\theta$ dependence of $\widetilde{\cal M}$ 
comes from higher partial waves with $J>J_0$.  
If the process has only one partial wave $J$ ({\it e.g.} for two-body 
decays or $e^+e^-\to \mu^+\mu^-$), the whole angular dependence can 
be extracted in the form of $d^J$ and the $\phi$-dependent phase factor. 

This technique can be applied to the process we consider if we note the 
following two points:

\item{1)} The amplitude can be decomposed into two parts: 
 (A) production of a virtual 
vector boson $V^*$, $e^+e^-\to V^*$ ($V=\gamma$ or $Z$); (B) the virtual 
vector boson decaying to $\zeta\bar\zeta g$.  The helicity amplitude 
for the whole process is the product of the two amplitudes with a fixed 
$V^*$ helicity ($\lambda_V=\pm1$, $0$), which are then summed over:  
$$ {\cal M}= - \mathop{\textstyle\sum}\limits_{V}{1\over s-m_V^2}
\mathop{\textstyle\sum}\limits_{\lambda_V=\pm1,0} 
{\cal M}(e^-e^+\to V^*)\,{\cal M}(V^*\to \zeta\bar\zeta g) \;.\Eq$$
(Note that we need three polarization states for both $\gamma^*$ and $Z^*$.  
A fourth (scalar) polarization state need not be included since $V^*$ 
couples to a conserved current.)

\item{2)} The helicity amplitude for the second $1\to 3$ process 
$V^* (q) \to \zeta (p) + \bar\zeta (\bar p) + g (k) $ can be treated 
in the same way as $2\to2$ processes if one works in the ``quasi-two-body'' 
frame, the c.m.\ frame 
of {\it two\/} of the final particles ({\it e.g.} $\zeta\bar\zeta$), 
not in the $V^*$ c.m.\ frame.  This frame is related to the $e^+e^-$ 
c.m.\ frame by a boost along the gluon direction.

First, we list the helicity 
amplitudes for $V^* (q,\lambda_V) \to \zeta (p) + \bar\zeta (\bar p) 
+ g (k,\lambda) $ (note that the scalar particles have zero helicities).  
We take the $V^*$ (and $g$) direction as the $z$ axis, and $\zeta$ 
direction as ($\theta$, $\phi$).  We follow the Jacob-Wick convention 
to fix the phase of the fermion/vector wave functions.  The amplitudes 
can be written as
$${\cal M}(V^*\to\zeta\bar\zeta g) = g_f g_s T^a \widetilde{\cal M}\;,\Eq$$
with 
$$g_f=\cases{ Q_\zeta e & for $V=\gamma$, \cr
            e (T_{3\zeta}-Q_\zeta\sin^2\!\theta_W)/\sin\theta_W \cos\theta_W
                 & for $V=Z$,\cr}\Eq$$
and $\widetilde{\cal M}$ given by the following:

\noindent $\lambda_n \equiv \lambda_V - \lambda = \pm2$ 
$$\widetilde{\cal M} = 
{2 \tau v^2 \sin^2\theta\over (1-\tau)(1-v^2\cos^2\theta)}
  e^{i\lambda_n\phi}\;,\Eq$$

\noindent $\lambda_n \equiv \lambda_V - \lambda = \pm1$ 
$$\widetilde{\cal M} = \mp {2\sqrt{2\tau} v^2 \sin\theta\cos\theta\over 
(1-\tau)(1-v^2\cos^2\theta)}  e^{i\lambda_n\phi}\;,\Eq$$

\noindent $\lambda_n \equiv \lambda_V - \lambda = 0$ 
$$\widetilde{\cal M} = 
-{2 \tau v^2 \sin^2\theta\over (1-\tau)(1-v^2\cos^2\theta)} - 2 \;.\Eq$$

The helicity amplitude for $e^- e^+ \to V^*$ is simpler.  We take the $V^*$ 
direction as the $z$ axis and the $e^-$ direction lying on the $xz$ plane 
(with a positive $x$ component).  The angle between the two directions 
is denoted by $\Theta$.  This choice is made to relate the two 
frames by the boost along the $z$ axis, under which the helicity of $V^*$ 
does not change.  The amplitude conserves electron chirality so that only 
$\lambda_i=\lambda(e^-) - \lambda(e^+) = \pm1$ is allowed.
$${\cal M}= (-1)^{\lambda_V} g_i \sqrt{2s}\, d^1_{\lambda_i\lambda_V}(\Theta) 
\;,$$
with
$$g_i=\cases{ -e & for $V=\gamma$, $\lambda_i=\pm1$,\cr
            e (v_e\mp a_e)/\sin\theta_W \cos\theta_W
                 & for $V=Z$, $\lambda_i=\pm1$.\cr}\Eq$$
The explicit form of the $d$ functions needed is
$$d^1_{\lambda\mu}(\Theta) = 
\cases{ {1\over2}(1+\lambda\mu\cos\Theta) & $|\lambda|=|\mu|=1$,\cr
-{\lambda\over\sqrt2}\sin\Theta & $|\lambda|=1$, $\mu=0$.\cr}\Eq$$

\Appendix{B}{Three-body phase space}

The differential cross section can be calculated from the helicity amplitudes 
with the Lorentz-invariant three-body phase space
\def\DP#1{{d^3#1\over (2\pi)^3 2 #1^0}}
$$ d\Phi_3 = (2\pi)^4\delta^4(q-p-\bar p-k)\,\DP{p}\DP{\bar p}\DP{k} \;,\Eq$$
which in terms of the kinematic variables defined in Section~3 is
$$ d\Phi_3 = {sv(1-\tau)\over 1024\pi^4}\,
d\tau\, d\cos\Theta\, d\cos\theta\, d\phi \;,\Eq$$
where $1-\beta^2<\tau<1$, and $v=(1-4m^2/\tau s)^{1/2}$ is the $\zeta$ 
velocity in the $\zeta\bar\zeta$ c.m.\ frame.  
(We have integrated over one azimuthal angle on which the cross section 
has trivial dependence.) 

Two of the kinematic variables, $\tau$ and $\theta$, specify the 
final state configuration and the other two determine the orientation 
of the configuration with respect to the initial beam axis.  
Integrating over $\Theta$ and $\phi$ one has
$$ d\Phi_3 = {sv(1-\tau)\over 256\pi^3}\, d\tau\, d\cos\theta  \;.\Eq$$

For analytic integration over the three-body phase space for $m\neq0$, 
it is easier to use ($\cos\theta$, $v$) as the integration 
variables, for which the integration region is $0<v<\beta$, 
$-1<\cos\theta<1$. 
$$ d\Phi_3 = {s(1-\beta^2)\over 128\pi^3}\,{v^2(\beta^2-v^2)\over(1-v^2)^3}
\, dv\,d\cos\theta \;.\Eq$$

The variables ($x$, $\bar x$) have a special property 
that the phase space factor is constant (Dalitz variables). 
$$ d\Phi_3 = {s\over 128\pi^3} dx\,d\bar x \;.\Eq$$
The translation is made using 
$$ \tau = 1-x-\bar x \;,\qquad v\cos\theta = -{x-\bar x\over x+\bar x}
\;.\Eq$$
The physical phase space region is bounded by the inequality
$$ x\bar x(1-x-\bar x) -{m^2\over s} (x+\bar x)^2 > 0 \;.\Eq$$

\Appendix{C}{Formulas for fermion pair production}

\REF\VectorC{B.~L. Ioffe, \PL78B, 277 (1978);\nextline 
G.~Grunberg, Y.~J. Ng, and S.-H.~H. Tye, \PRD21, 62 (1980); 
{\bf 36}, 311(E) (1987).}
\REF\AxialC{%
J.~Jers\'ak, E.~Laermann, and P.~M. Zerwas, \PRD25, 1218 (1982);\nextline
A.~Djouadi, \ZPC39, 561 (1988);\nextline 
J.~G. K\"orner, A.~Pilaftsis, and M.~M. Tung, \ZPC63, 575 (1994).}

\begingroup
\def\+{{+}}
\def\-{{-}}
We collect results for the familiar process of fermion pair production 
$e^+e^-\to F\bar F(g)$\refmark{\Schwinger,\VectorC,\AxialC} in this 
Appendix.  These results have been independently calculated by us.  
The $ F\bar F\gamma$ and $F\bar FZ$ vertices in the lowest order have the 
form
$$ -ieQ_F \gamma_\mu \quad\hbox{and}\quad -{ie\over\cos\theta_W\sin\theta_W}
 \gamma_\mu(v_F - a_F\gamma_5) \;,\Eq$$
with
$$\eqalignno{
v_F&={1\over2}\bigl[T_{3}(F_L)+T_{3}(F_R)\bigr]-Q_F\sin^2\!\theta_W \;,
&\Eq\cr\r
a_F&= {1\over2}\bigl[T_{3}(F_L)-T_{3}(F_R)\bigr] \;.&\Eq\cr}$$
At ${\cal O}(\alpha_s)$, the correction factor is common for the photon and 
the vector part of the $Z$ current.  The axial part of the $Z$ current 
is modified differently if $m_F\neq0$.  We write the general Lorentz 
structure of the vector and axial vertices for on-shell fermions
$$\Gamma_\mu = F_1^V(q^2) \gamma_\mu + F_2^V(q^2) i\sigma_{\mu\nu}q^\nu/m_F 
\Eq$$
for the vector current and
$$\Gamma_\mu^5 = F_1^A(q^2) \gamma_\mu\gamma_5 + F_2^A(q^2) q_\mu \gamma_5 
/m_F \Eq$$
for the axial current.  
We have retained only terms appearing at ${\cal O}(\alpha_s)$, {\it i.e.}, 
terms allowed by $CP$ invariance and vector current conservation.
We normalize the vertices such that $F_1^{V,A}=1$ at the lowest order.   
$F_2^{V,A}$ do not appear in this order.  

We write the ${\cal O}(\alpha_s)$ corrected vertices as 
$$\eqalignno{F_1^{V,A}(q^2) &= 1 + {C_{\!R}\alpha_s\over 2\pi} f_1^{V,A}(q^2)
\;,&\Eq\cr\r
F_2^{V,A}(q^2) &=  {C_{\!R}\alpha_s\over 2\pi} f_2^{V,A}(q^2)
\;.&\Eq\cr}$$
We find the renormalized vector form factors for $q^2>4m_F^2$
$$\eqalignno{
f_1^V &=  
\biggl( - {1+\beta^2\over 2\beta} \log{1\+\beta\over 1\-\beta}+1\biggr) 
\log{m_F^2\over\lambda^2}\cr\r
&\quad +{1+\beta^2\over\beta}\biggl[  \Li2({1\-\beta\over1\+\beta}) 
+  \log{1\+\beta\over2\beta} \log{1\+\beta\over 1\-\beta}
- {1\over4} \log^2{1\+\beta\over1\-\beta} + {\pi^2\over3} \biggr]\cr\r
&\quad +{1+2\beta^2\over2\beta}\log{1\+\beta\over1\-\beta} - 2 \cr\r
&\quad + i\pi\biggl[ {1+\beta^2\over 2\beta}\biggl( 
\log{m_F^2\over\lambda^2} + \log{4\beta^2\over1\-\beta^2}\biggr) 
-{1+2\beta^2\over 2\beta}\biggr] \;,&\Eq\cr\r
f_2^V &= {1-\beta^2\over4\beta} 
\biggl(-\log{1\+\beta\over 1\-\beta}+i\pi\biggr) \;,&\Eq\cr}$$
where $ \beta = (1-4m_F^2/ q^2)^{1/2} $ .  We have used on-mass-shell 
renormalization condition such that $F_1^V(0)=1$.  Once the prescription 
for the vector vertex is fixed, there is no freedom to choose a 
renormalization condition for the axial vertex because of electroweak 
gauge symmetry.  In particular, $F_1^A(0)$ {\it deviates\/} from unity.  
The axial form factors are found to be
$$\eqalignno{
f_1^A &= 
\biggl( - {1+\beta^2\over 2\beta} \log{1\+\beta\over 1\-\beta}+1\biggr) 
\log{m_F^2\over\lambda^2}\cr\r
&\quad +{1+\beta^2\over\beta}\biggl[  \Li2({1\-\beta\over1\+\beta}) 
+  \log{1\+\beta\over2\beta} \log{1\+\beta\over 1\-\beta}
- {1\over4} \log^2{1\+\beta\over1\-\beta} + {\pi^2\over3} \biggr]\cr\r
&\quad +{2+\beta^2\over2\beta}\log{1\+\beta\over1\-\beta} - 2 \cr\r
&\quad + i\pi\biggl[ {1+\beta^2\over 2\beta}\biggl( 
\log{m_F^2\over\lambda^2} + \log{4\beta^2\over1\-\beta^2}\biggr) 
-{2+\beta^2\over 2\beta}\biggr] \;,&\Eq\cr\r
f_2^A &= {1-\beta^2\over 4\beta} \biggl[ (2+\beta^2) 
\biggl( -\log{1\+\beta\over 1\-\beta}+i\pi\biggr) + 2\beta \biggr] \;.&\Eq\cr}
$$
The second form factor $F_2^A$ represents the pseudoscalar part of the 
axial current and does not contribute to the reaction we are interested in 
as long as the electron mass is neglected.

The differential cross section for the 3-body $F\bar Fg$ final state 
is for the vector current
$$\eqalignno{{d\sigma^V_\pm\over d\tau\,d\cos\theta} 
&= \sigma_{0\pm}^V\,{C_{\!R}\alpha_s\over\pi}\,
{v(1-\tau)\over \beta(3-\beta^2)}\,\cr\r
&\quad\times
\biggl[ {1+v^2\cos^2\!\theta\over 1-v^2\cos^2\!\theta}
+ {2(3-\beta^2)\tau v^2\sin^2\!\theta\over
(1-\tau)^2 (1-v^2\cos^2\!\theta)^2 }\biggr]\;,&\Eq\cr}$$
and for the axial current
$$\eqalignno{{d\sigma_\pm^A\over d\tau\,d\cos\theta}
&= \sigma_{0\pm}^A\,{C_{\!R}\alpha_s\over\pi}\,
{v(1-\tau)\over 2\beta^3}\,\cr\r
&\quad\times
\biggl[ {2-\beta^2+v^2\cos^2\!\theta\over 1-v^2\cos^2\!\theta}
+ {4\beta^2\tau v^2\sin^2\!\theta\over
(1-\tau)^2 (1-v^2\cos^2\!\theta)^2 }\biggr]\;,&\Eq\cr}$$
with the lowest order cross sections
$$\eqalignno{
\sigma_{0\pm}^V &= {8\pi d_R\alpha^2\over 3s}\,H_{V\pm}^{2} 
{\beta(3-\beta^2)\over 2} \;,&\Eq\cr\r
\sigma_{0\pm}^A &= {8\pi d_R\alpha^2\over 3s}\,H_{A\pm}^{2} 
{\beta^3} \;.&\Eq\cr}$$
Here $H_{V\pm}$ and $H_{A\pm}$ are given by
$$\eqalignno{
H_{V\pm} &= - Q_F 
+ {(v_e {\mp} a_e) v_F  
\over \cos^2\!\theta_W \sin^2\!\theta_W}{s\over s-m_Z^2} \;,&\Eq\cr\r
H_{A\pm} &=  {(v_e {\mp} a_e) a_F  
\over \cos^2\!\theta_W \sin^2\!\theta_W}{s\over s-m_Z^2} \;.&\Eq\cr}$$

In terms of the Dalitz variables, we have for the vector current
$$\eqalignno{{d\sigma_\pm^V\over dx\,d\bar x} 
&= \sigma_{0\pm}^V\,{C_{\!R}\alpha_s\over\pi}\,{1\over\beta}\,
\biggl[ {1\over 3-\beta^2} \biggl({\bar x\over x}+{x\over\bar x}\biggr) \cr\r
&\qquad\qquad\qquad - \left({1\over x} + {1\over \bar x} \right)
+ {1+\beta^2\over 2}\,{1\over x\bar x} 
-{1-\beta^2\over 4} \,\left({1\over x^2}+{1\over \bar x^2}\right)\biggr]
\;,\qquad&\Eq\cr}$$
and for the axial vector current
$$\eqalignno{{d\sigma_\pm^A\over dx\,d\bar x} 
&= \sigma_{0\pm}^A\,{C_{\!R}\alpha_s\over\pi}\,{1\over\beta}\,
\biggl[ {3-\beta^2\over 4\beta^2} \biggl({\bar x\over x}+{x\over\bar x}\biggr)
+ {1-\beta^2 \over 2\beta^2} \cr\r
&\qquad\qquad\qquad - \left( {1\over x} + {1\over \bar x}\right)
+ {1+\beta^2\over 2}\,{1\over x\bar x} 
-{1-\beta^2\over 4} \,\left({1\over x^2}+{1\over \bar x^2}\right)\biggr]
\;.\qquad&\Eq\cr}$$

The total ${\cal O}(\alpha_s)$ correction can be written as the sum of 
the three contributions, virtual, soft, and hard corrections.  We list each 
correction term for the scalar-pair production for comparison 
$$\eqalignno{
\Delta_{\rm virtual} &= {1\over\beta} A_v(\beta) 
+ {1+\beta^2\over\beta} \log{1+\beta\over 1-\beta} - 2 \;,&\Eq\cr\r
\Delta_{\rm soft} &= {1\over\beta} A_s(\beta) 
+ {1\over\beta}\log{1+\beta\over 1-\beta} \;,&\Eq\cr\r
\Delta_{\rm hard} &= {1\over\beta} A_h(\beta) 
- {1\over 4\beta^3}(3+\beta^2)(1-\beta^2) \log{1+\beta\over 1-\beta} 
+ {1\over 2\beta^2}(3+7\beta^2) \;,\qquad&\Eq\cr}$$
where we have collected the ``dilogarithmic'' part (dilogarithm and 
double log terms) into the functions $A_i$ ($i=v$, $s$, $h$)
$$\eqalignno{
A_v(\beta) &= \biggl(-{1\+\beta^2\over 2} 
\log{1\+\beta\over 1\-\beta} + \beta \biggr) \log{m^2\over\lambda^2} \cr\r
&\quad + (1 +\beta^2) \biggl[ \Li2(1\-\beta\over 1\+\beta) 
- \log{2\beta\over 1\+\beta}\log{1\+\beta\over 1\-\beta} 
- {1\over4}\log^2{1\+\beta\over1\-\beta} + {\pi^2\over3} \biggr] \;,\cr
&&\Eq\cr
A_s(\beta) &= \biggl({1\+\beta^2\over2} \log{1\+\beta\over 1\-\beta} 
  -\beta\biggr) \log{4\omega^2\over\lambda^2} \cr\r\noalign{\goodbreak} 
&\quad  +(1 +\beta^2)\biggl[\Li2({1\-\beta\over 1\+\beta}) 
         - \log{2\beta\over 1\+\beta}\log{1\+\beta\over 1\-\beta} 
           -{1\over4}\log^2{1\+\beta\over 1\-\beta} - {\pi^2\over6} \biggl]
\;,\cr &&\Eq\cr 
A_h(\beta) &= \biggl({1\+\beta^2\over2} \log{1\+\beta\over 1\-\beta} 
 -\beta\biggr) \log{m^2\over4\omega^2} \cr\r 
&\quad +(1 +\beta^2) \biggl[ 2\,\Li2({1\-\beta\over 1\+\beta})
         + 2\,\Li2(-{1\-\beta\over 1\+\beta})
          -\log{2\over1\+\beta}\log{1\+\beta\over 1\-\beta} \cr\r
&\qquad\qquad\qquad 
+ {1\over2} \log^2{1\+\beta\over 1\-\beta} - {\pi^2\over6} \biggr]
 - 3\beta\log{4\over 1 \- \beta^2}  - 4\beta \log\beta\;.&\Eq\cr}$$
The function $A(\beta)$ in \eqA\ is the sum of these three
$$ A(\beta) = A_v(\beta) + A_s(\beta) + A_h(\beta) \;.\Eq$$

Turning to the fermion-pair production, it is found that the dilogarithmic 
terms are exactly the same as for the scalar-pair production.  For the vector 
part we find
$$\eqalignno{
\Delta^V_{\rm virtual} &= \Re f_1^V + {6\over 3-\beta^2} \Re f_2^V \cr\r
&= {1\over\beta} A_v(\beta) 
+ {\beta(4-\beta^2)\over 3-\beta^2} \log{1+\beta\over 1-\beta} - 2 \;,&\Eq\cr\r
\Delta^V_{\rm soft} &= {1\over\beta} A_s(\beta) 
+ {1\over\beta}\log{1+\beta\over 1-\beta} \;,&\Eq\cr\r
\Delta^V_{\rm hard} &= {1\over\beta} A_h(\beta) 
+ {9-2\beta^2+\beta^4\over 8\beta(3-\beta^2)}\log{1+\beta\over 1-\beta} 
+ {39-17\beta^2\over 4(3-\beta^2)} \;,&\Eq\cr}$$
and for the axial part
$$\eqalignno{
\Delta^A_{\rm virtual} &= {1\over\beta} A_v(\beta) 
+ {2+\beta^2\over 2\beta} \log{1+\beta\over 1-\beta} - 2 \;,&\Eq\cr\r
\Delta^A_{\rm soft} &= {1\over\beta} A_s(\beta) 
+ {1\over\beta}\log{1+\beta\over 1-\beta} \;,&\Eq\cr\r
\Delta^A_{\rm hard} &= {1\over\beta} A_h(\beta) 
+ {1\over 32\beta^3}(21-5\beta^2+3\beta^4-3\beta^6)\log{1+\beta\over 1-\beta} 
\cr\r &\quad + {1\over 16\beta^2}(-21+62\beta^2+3\beta^4) \;.&\Eq\cr}$$

\endgroup

\par \penalty-400 \vskip\chapterskip
   \spacecheck\referenceminspace \immediate\closeout\referencewrite
   \referenceopenfalse
   \leftline{\twelvebold References}\par
   \nobreak\vskip\headskip \penalty 30000
   \input reference.aux
   \endpage
\nopagenumbers
\FIG\figone{Feynman diagram for $e^+e^-\to\zeta\bar\zeta$ at the lowest 
order.}
\FIG\figtwo{Feynman diagrams for $e^+e^-\to\zeta\bar\zeta$ at 
${\cal O}(\alpha_s)$.}
\FIG\figthree{Feynman diagrams for $e^+e^-\to\zeta\bar\zeta g$.}
\FIG\figfour{Total ${\cal O}(\alpha_s)$ correction to the pair production 
cross section of scalar pair (solid); fermion pair via vector current (dash);
fermion pair via axial current (dotted).  All curves are for particles 
in the fundamental color representation ($C_{\!R}=4/3$) and the strong 
coupling constant $\alpha_s=0.12$.}
\FIG\figfive{Three-jet fraction for scalar (solid) and fermion (dashed) 
at the high energy limit, for $\alpha_s=0.12$ and $C_{\!R}=4/3$.}
\FIG\figsix{Three-jet fraction for scalar (solid), fermion-vector (dashed), 
and fermion-axial (dotted) for $\beta=0.4$, 0.6, 0.8, and 1.0$, 
\alpha_s=0.12$ and $C_{\!R}=4/3$.}
\FIG\figseven{Same as Fig.~\figsix, but as a function of a jet-defining 
variable $y_c$.}
\par \penalty-400 \vskip\chapterskip
   \spacecheck\referenceminspace \immediate\closeout\figurewrite
   \figureopenfalse
   \leftline{\twelvebold Figure Captions}\par
   \nobreak\vskip\headskip \penalty 30000
   \input figures.aux
   
\endpage
\vfil
$$\epsfbox{fig1.eps}$$
\endpage
\epsfysize=550pt
$$\epsfbox{fig2.eps}$$
\endpage
$$\epsfbox{fig3.eps}$$
\endpage
\epsfxsize=400pt
$$\epsfbox{fig4.eps}$$
\endpage
\epsfxsize=400pt
$$\epsfbox{fig5.eps}$$
\endpage
\epsfxsize=400pt
$$\epsfbox{fig6.eps}$$
\endpage
\epsfxsize=400pt
$$\epsfbox{fig7.eps}$$
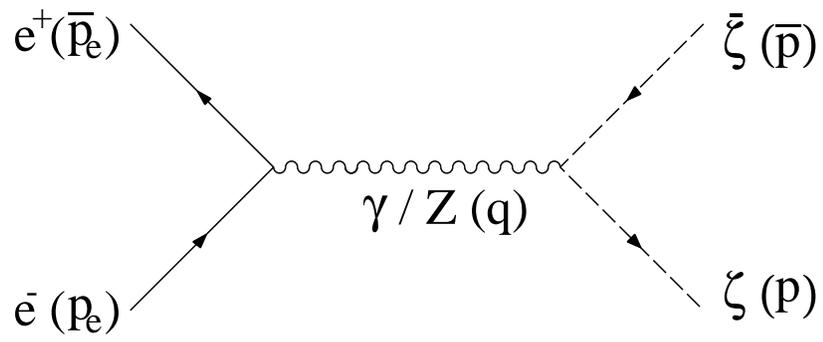
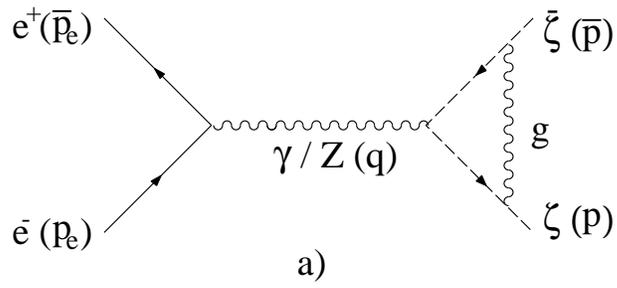
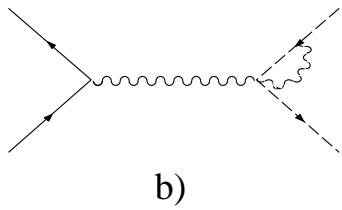
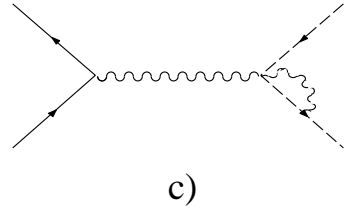
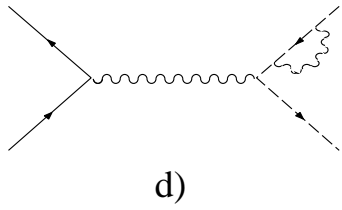
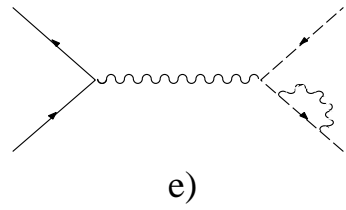
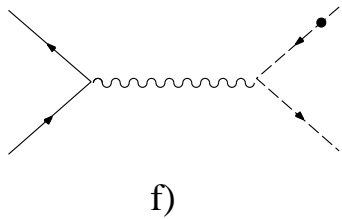
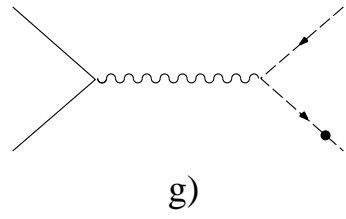
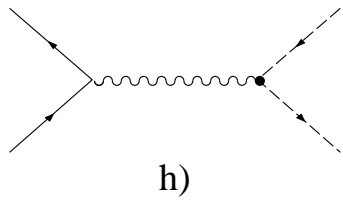
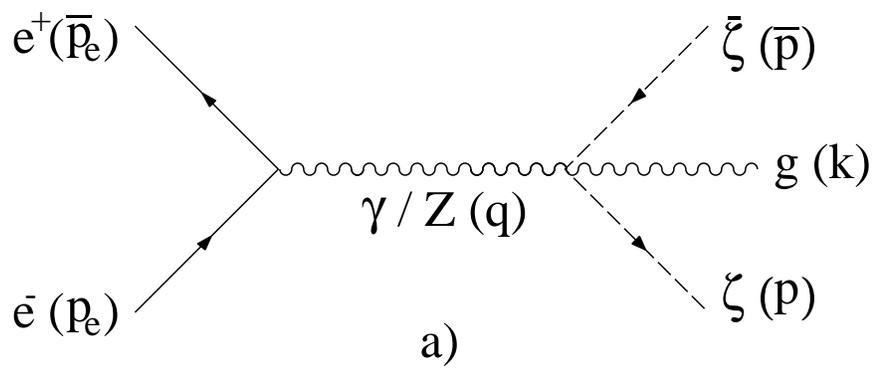
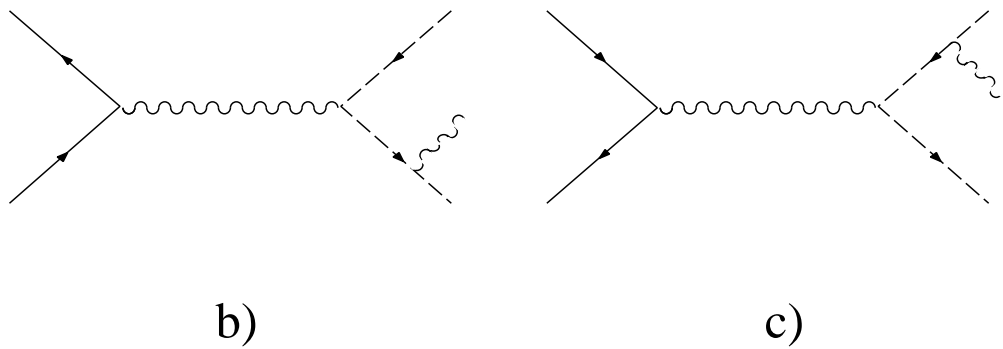
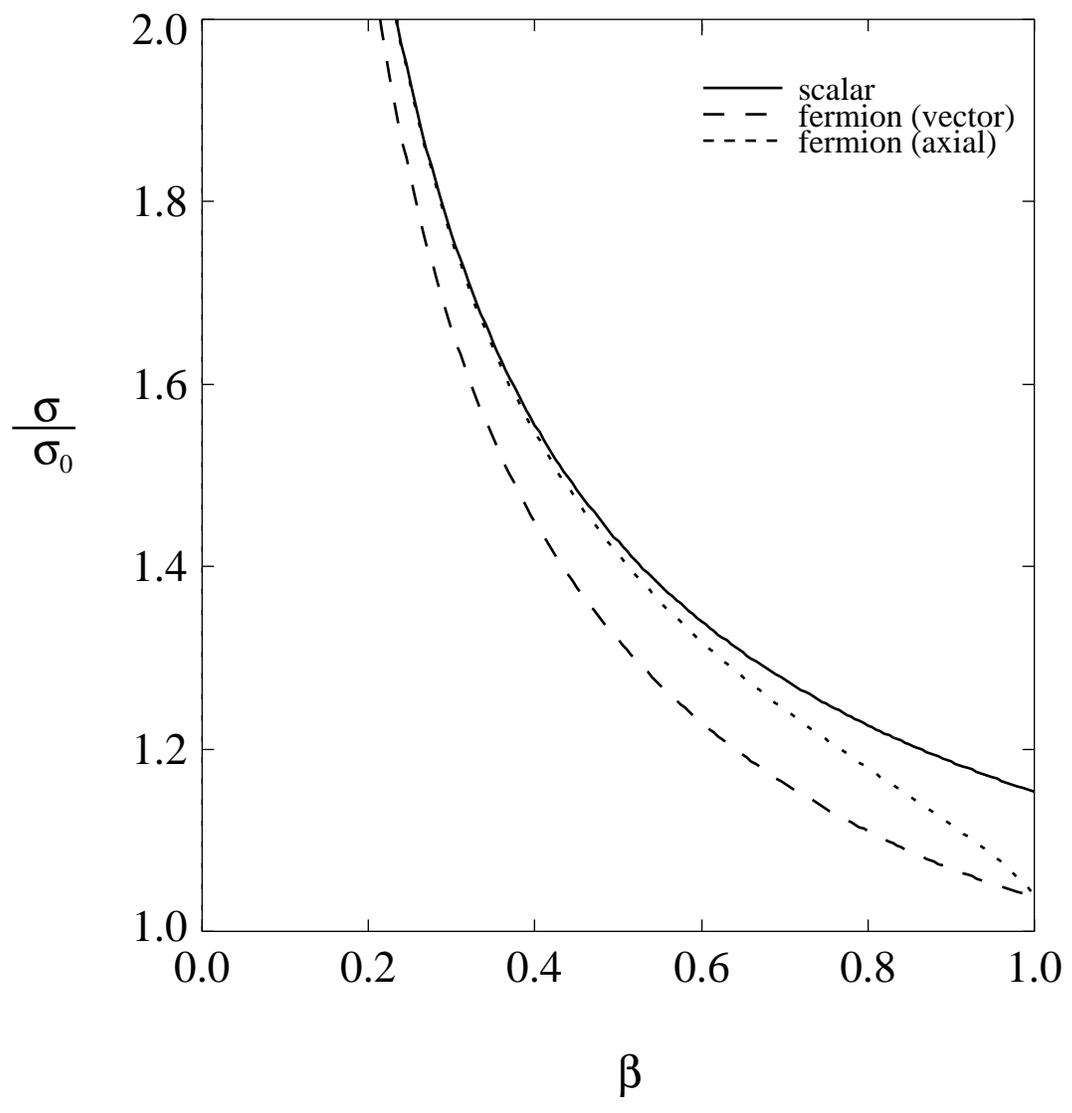
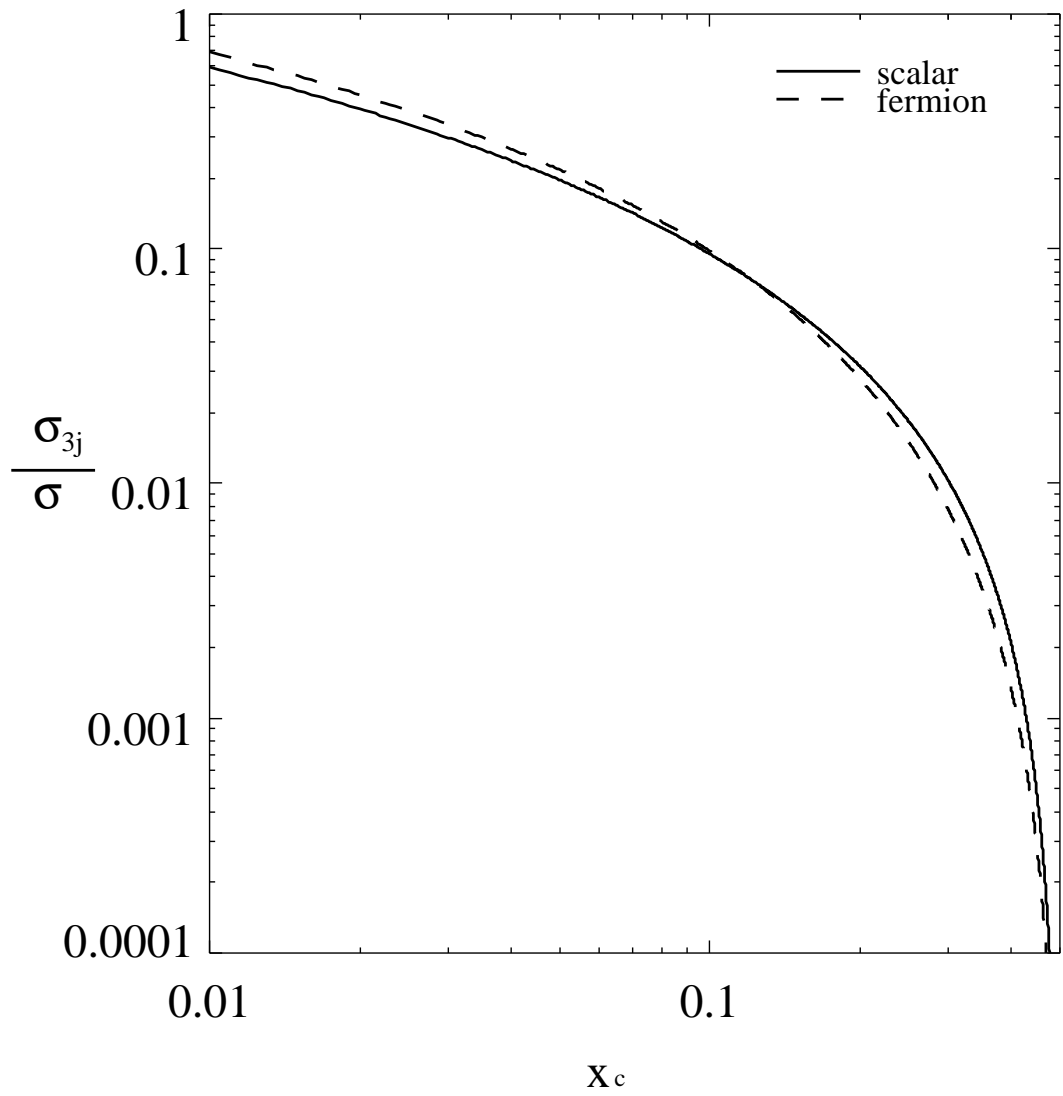
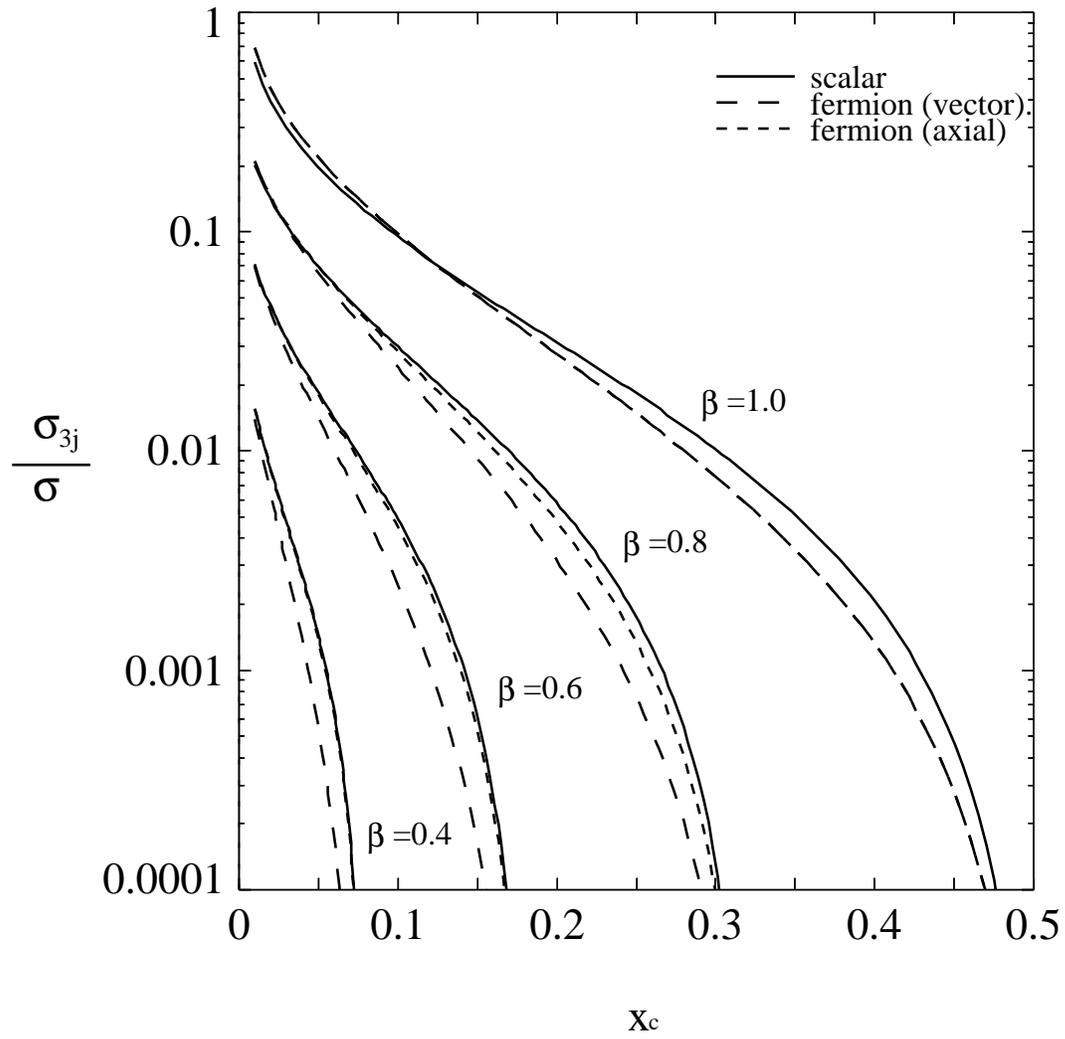
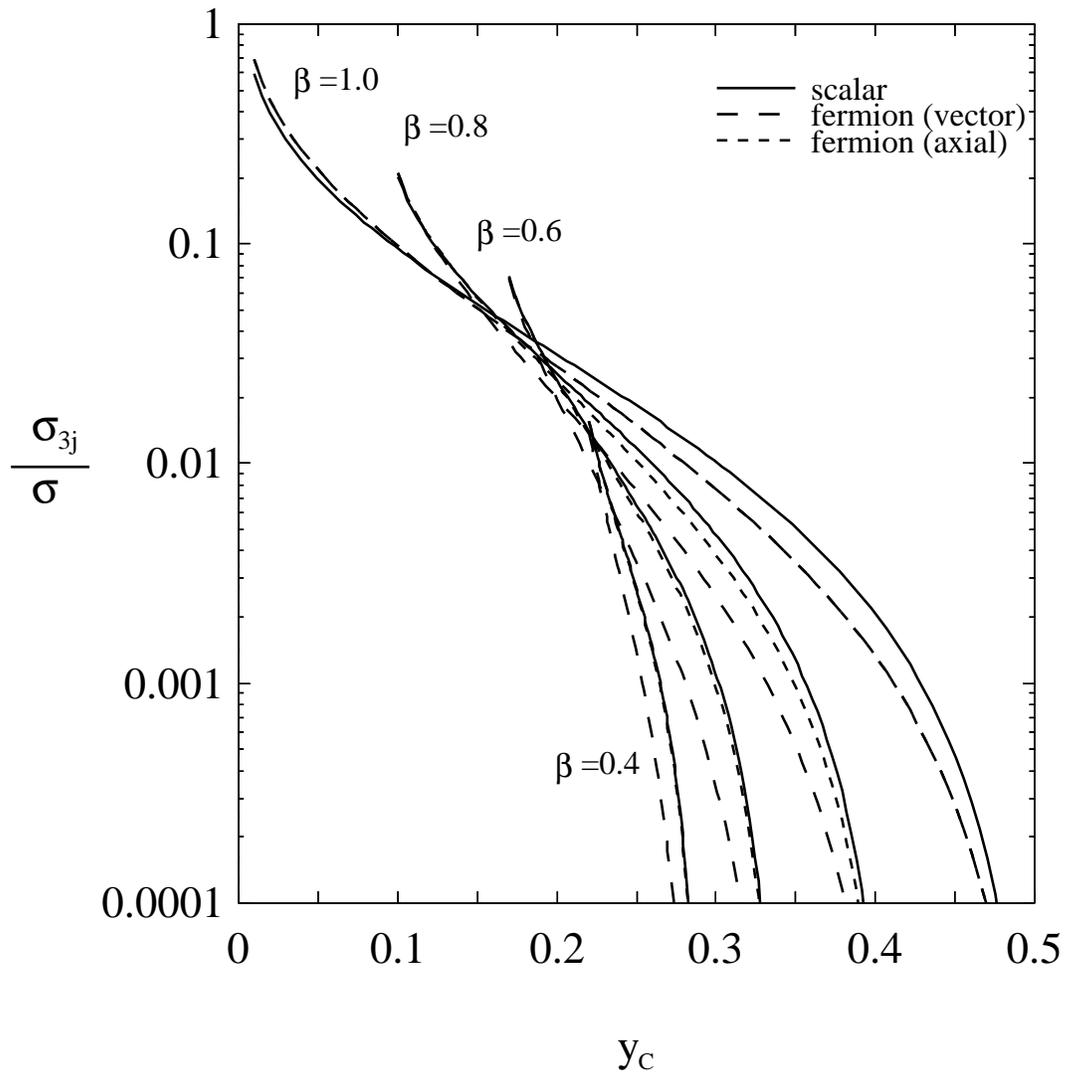
\bye